QIS-XML: An Extensible Markup Language
for Quantum Information Science

A thesis submitted in partial fulfillment of the requirements for the degree of Master of Science at George Mason University

By

Pascal Heus
Gradué en informatique
Institut Paul Lambin, Université Catholique de Louvain (UCL), 1985

Director: Richard Gomez
College of Science, Computational and Data Sciences

Spring Semester 2011
George Mason University
Fairfax, VA

# ACKNOWLEDGEMENTS

I would like to thank Dr. Richard Gomez for introducing me to Quantum Information Science, teaching me the underlying principles, and guiding me throughout my research.

I'm grateful to Dr. Blaisten-Bajoras for encouraging and supporting me towards the pursue of my Master degree.

Endless thanks to my family for supporting me throughout my (many) years of study.

# TABLE OF CONTENTS

# LIST OF FIGURES

# LIST OF ABBREVIATIONS

CIS           Classic Information Science
ICT           Information & Communication Technology
ISO           International Standard Organization
IT             Information Technology
QC           Quantum Computing
QIS           Quantum Information Science
QIT           Quantum Information Technology
SOA          Service Oriented Architecture
SOAP        Simple Object Access Protocol
SVG         Scalable Vector Graphics
XSL          Extensible Stylesheet Language
XSLT        XSL Transformation
XML         Extensible Markup Language
W3C         World Wide Web consortium

# ABSTRACT

ADOPTION CHALLENGES FOR QUANTUM INFORMATION TECHNOLOGY

Pascal Heus, MS

George Mason University, 2011

Thesis Director: Dr. Richard Gomez

This project examines issues of interoperability and integration between the Classic Information Science (CIS) and Quantum Information Science (QIS). This paper provides a short introduction to the Extensible Markup Language (XML) and proceeds to describe the development steps that have lead to a prototype XML specification for quantum computing (QIS-XML). QIS-XML is a proposed framework, based on the widely used standard (XML) to describe, visualize, exchange and process quantum gates and quantum circuits. It also provides a potential approach to a generic programming language for quantum computers through the concept of XML driven compilers. Examples are provided for the description of commonly used quantum gates and circuits, accompanied with tools to visualize them in standard web browsers. An algorithmic example is also presented, performing a simple addition operation with quantum circuits and running the program on a quantum computer simulator.

Overall, this initial effort demonstrates how XML technologies could be at the core of the architecture for describing and programming quantum computers. By leveraging a widely accepted standard, QIS-XML also builds a bridge between classic and quantum IT, which could foster the acceptance of QIS by the ICT community and facilitate the understanding of quantum technology by IT experts. This would support the consolidation of Classic Information Science and Quantum Information Science into a Complete Information Science, a challenge that could be referred to as the "Information Science Grand Unification Challenge".

# 1  BACKGROUND

## 1.1  INFORMATION TECHNOLOGY, METADATA AND XML

### 1.1.1  What is Metadata?

In the ICT world, information collected on objects or data is commonly referred to as "Metadata". While many definitions exist for this term, the general consensus is to say that Metadata is "Data bout Data". Metadata does not change anything about the item it describes but rather defines its nature but attaching a set of descriptive attributes to the object. Simple examples are the title of a book, the brand or the color of a car, the name of a person, the number of calories of a food item, etc. This is further illustrated by the figure below (Adobe Systems Incorporated n.d.) where we can see on the left cans without any metadata and on the right the same items with metadata.

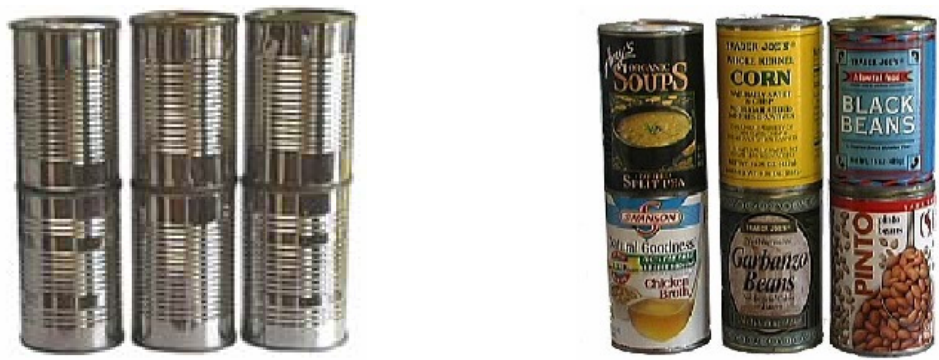

*Figure 1 - Objects without and with metadata*

An important aspect of metadata is that it can be stored, accessed or exchanged without having to pass along the underlying objects. We can browse a library catalog remotely or select a car from a brochure with going to the dealership.

In a similar way, we could describe quantum hardware, gates, circuits or algorithms without having actual access to a real quantum computer, whether it exists or not.

### 1.1.2 What is XML?

Nowadays, information systems commonly capture metadata using a format called the Extensible Markup Language or XML. It is a system that "tags" or "markups" elements of information and stores them in a standard text format that can be read by any computer. XML is broadly used on the Internet, in particular to exchange information between organizations.  It has emerged in recent years as a dominant technology. While invisible to most end users, XML is driving todays' Web.

XML is actually a term that encompasses several technologies and functionalities. Adopting XML as a common language allows information systems not only to capture metadata (in XML) but also to validate it against agreed upon specifications (using DTD or XSchema), transform it to other formats such as HTML or PDF (using XSLT), search it to build catalogs and lookup information (using XPath or XQuery), exchange it (using SOAP based web services) and even edit it (using XForms). All these functionalities are inherent to the XML technology and require little efforts to implement.

Here are simple illustrations of what XML documents look like:

```
<catalog>                                  <mycontacts>
   <book isbn="0385504209">                   <person>
      <title>Da Vinci Code</title>               <name>John Doe</name>
      <author>Dan Brown</author>                 <email>jdoe@example.org</email>
   </book>                                     </person>
   <book isbn="0553294385">                   <person>
      <title>I, robot</title>                    <name>Richard Feynman<name>
      <author>Isaac Asimov</author>
   </book>                                  <occupation>scientist</occupation>
</catalog>                                     </person>
                                           </mycontacts>
```

*Figure 2 – Simple examples of XML*

The language seems and is pretty simple and can be read by users. XML is not a proprietary format and is stored as simple text. We can also see that <elements> of information are delimited by an "<" opening and a ">" closing "tag" than contains text or other elements (children). In some case, "attributes" can also be directly embedded in the tag itself, as it is the case for the ISBN number of the book. While the syntax of XML contains a few more rules, this provides a general idea of its basic principles. What is important to understand is that it can be used to describe many different structures: it is extensible.

### 1.1.3   Different metadata, different specifications

Having a common language to describe objects is a start but is not enough to satisfy all metadata management needs. We also need a way to define and *structure* the elements of information we want to use when describing an object. For example, we would not describe a book or a car using the same attributes. Even if we are talking about the same object, it is very easy to come up with different structures or attributes.

If we want to be able to universally manage and exchange metadata, we need to agree on *common structures* of descriptive elements to describe the various objects. This is what is known in XML as a *specification* and is implemented using the DTD or XSchema languages mentioned above. The XML world is made of hundreds of such specifications, each specializing in a specific type of object such as books, press releases, cars, stock quotes, etc. It is not uncommon for several specifications to exist for the same type of object and they sometime compete with each other. It is then up to each agency to select one or to implement several at the same time (it is also actually quite easy to transform one specification into another). XML specifications can also be submitted to the International Standard Organization (ISO) to become an official ISO standard.

# 2 RESEARCH OVERVIEW

## 2.1 QUANTUM COMPUTING AND METADATA

As an emerging field of study, QIS currently lacks an XML specification (or schema). No structure has been defined to describe the "objects" or the "concepts" surrounding QIS. This project is an initial effort to define an XML specification to describe objects such as:

- Qubits and unitary transformations;

- Single qubit quantum gates along with their breakdown into universal gates;

- Multiple qubit quantum gates;

- Quantum circuits;

- Quantum algorithms;

### 2.1.1 Methods

To design the QIS-XML specification, I used common XML tools such as Altova XMLSpy and Oxygen, specialized editors and development environment for modeling, editing, debugging, and transforming all XML technologies. The software was also used to generate the QIS-XML diagrams and examples presented throughout this document. A traditional C++ complier was used to develop and customize utility software.

## 2.2  OBJECTIVES

Having a metadata structure in place will greatly facilitate the management and exchange of information by the QIS research community.

I also see this as the foundation to:

- Build and share knowledge between experts and make it more accessible to the general public

- Develop educational tools

- Outline a basic programming language for quantum computers

- Design a rudimentary compiler to transform programming instructions into a quantum circuit and universal quantum gates (an assembly language for quantum computing)

- Implement tools that will assemble circuits or program circuits to execute the algorithms

Using an XML based approach ensures that the framework fits in the classic ICT environment which will in turn facilitate adoption and interaction between classic and quantum based systems.

# 3   THE QIS-XML MODEL

Defining an XML schema is often referred to as "modeling". A fairly close parallel can actually be drawn between an XML schema and the Unified Modeling Language (UML). The process consists in describing the hierarchical set of elements, attributes and rules that define and regulate the information structure. Note that as the English dictionary is written in English, an XML schema is actually written in the XML language as well (a schema is an XML document, usually with a .xsd extension).

Once a schema has been created, an XML "document" can be associated with it (through what is called a "namespace") and "validated" against it. This validation process ensures that the document complies with the structure and rules defined by the schema.

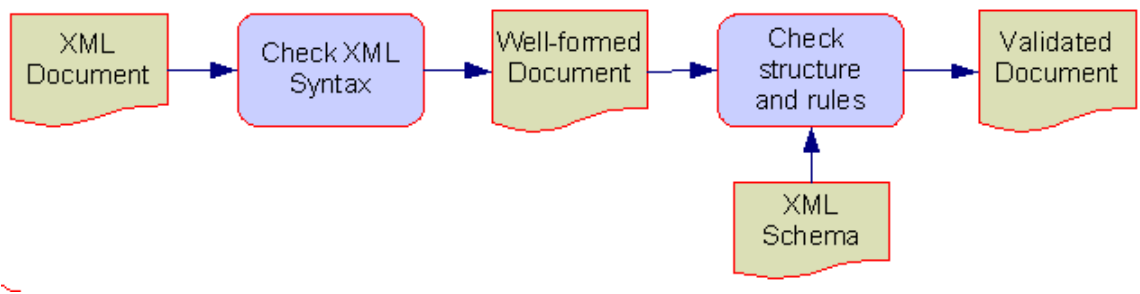

*Figure 3 - XML document validation process*

## 3.1  TOP LEVEL STRUCTURE AND SPECIAL DATA TYPES

By rule, an XML document can only have one root element. I defined the <QIS> element as top-level container for three types of objects: Gate, Circuit and Algorithm as illustrated below.

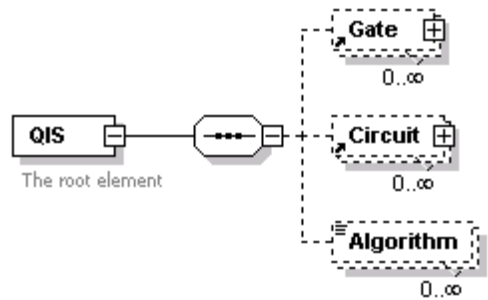

*Figure 4 - A QIS documents is a sequence of 0 or more Gate, Circuit or Algorithm elements*

As XML is only aware of common data types, the initial task before proceeding into the core elements of information was to create new elementary structures.

### 3.1.1  Complex Number

XML does not come with a complex number data type. I therefore created an element that can hold a *real* and *imaginary* value.

Later on, as I started to practically describe gates, I quickly realized that a quantitative value is sometimes not enough and it is useful to also have the option to provide a *symbolic* expression that describe the complex number. As this symbolic expression can be software or environment dependent, it can be repeated as often as necessary.

I therefore defined a <ComplexNumber> as an element with two attributes - @r and @i - to capture quantitative real and imaginary values along with a sequence of symbolic expressions.

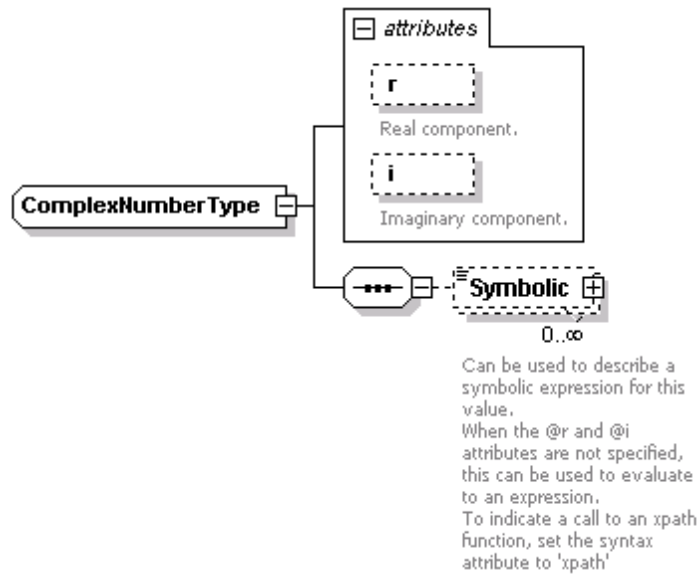

*Figure 5 - The Complex Number element type*

### 3.1.2   Matrix

I also had to define a couple of structures to manage matrices. I first defined the <Matrix> itself as a *sequence of cells*. A <Matrix> comes with two attributes – @rows and @cols – to specify its dimensions.

A <MatrixCell> is then defined as a <ComplexNumber> with two extra attributes – @row and @col – to specify the location of the cell within the matrix.

I also made the general assumption that if a cell is not defined within a matrix, its value is zero.

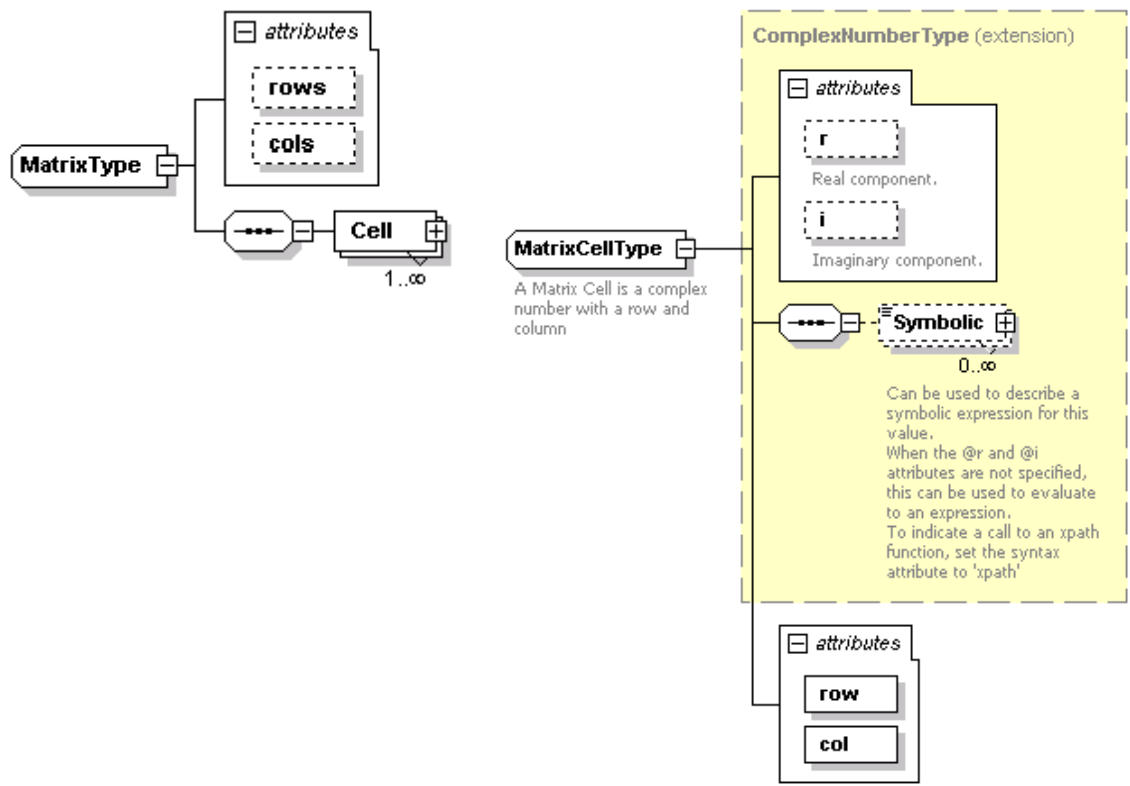

*Figure 6 – The Matrix and MatrixCell types*

### 3.1.3 Elementary components

With the basic data types in place, I proceeded to define two QIS specific elementary components: qubit and unitary transformation.

### 3.1.3.1 Qubit

The <Qubit> type is simply defined as a collection of two complex numbers: zero and one.

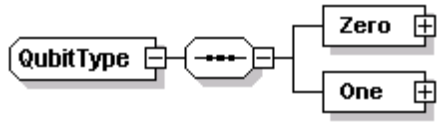

*Figure 7 - The Qubit element type*

### 3.1.3.2 Unitary Transformation

I defined the <UnitaryTransformation> as a special type of matrix: instead of having rows and columns, it is always square and has a @size attributes that specifies the number of qubit of the transformation.

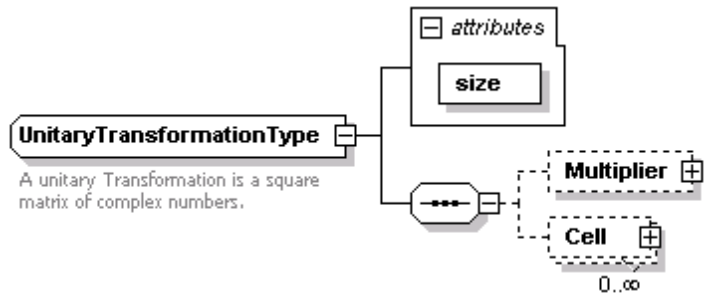

*Figure 8 - The UnitaryTransformation element type*

A size 1 is therefore a 2x2 matrix, size 2 a 4x4, etc. I quickly found it useful to add a Multiplier element as a complex number to scale the whole matrix (as in the Hadamard gate). A couple of XML examples of transformations are shown below.

| Unitary Transformation for Toffoli Gate | Unitary Transformation for Hadamard Gate |
|---|---|
| ```<br><Transformation size="3"><br>    <Cell row="1" col="1" r="1"/><br>    <Cell row="2" col="3" r="1"/><br>    <Cell row="3" col="2" r="1"/><br>    <Cell row="4" col="4" r="1"/><br>    <Cell row="5" col="5" r="1"/><br>    <Cell row="6" col="6" r="1"/><br>    <Cell row="7" col="8" r="1"/><br>    <Cell row="8" col="7" r="1"/><br></Transformation><br>``` | ```<br><Transformation size="1"><br>    <Multiplier r="0.707106781"><br>        <Symbolic syntax="odf">1/sqrt(2)</Symbolic><br>        <Symbolic<br>syntax="html">1/sqrt(2)</Symbolic><br>    </Multiplier><br>    <Cell row="1" col="1" r="1"/><br>    <Cell row="1" col="2" r="1"/><br>    <Cell row="2" col="1" r="1"/><br>    <Cell row="2" col="2" r="-1"/><br></Transformation><br>``` |

*Figure 9 - Examples of unitary transformations*

## 3.2  QUANTUM GATES

With the above data types available, I then started to describe the quantum <Gate> element. Fundamentally, a quantum gate can be simply defined as a <UnitaryTransformation>. It is therefore a required element of the <Gate> structure. I however wanted to provide further description of the <Gate> and added a few elements such as a <Name>, <Nickname> and <Description>. Since I expected a <Gate> to be reused elsewhere (like in <Circuit>), I also included a <Identification> element to be able to refer to it.

One important functionality that I wanted to associate with the <Gate> concept was the ability to describe "Equivalent Circuits". This is motivated by the idea that quantum gates can be represented in terms of sets of universal gates. For example, the set {Hadamard, Phase, C-NOT, pi/8} gates can be used to construct any quantum operations (Nielsen and Chuang 2000). Other example includes the 3-qubit Deutsch Gate or even universal circuit (Sousa and Ramos 2006). I therefore added an <EquivalentCircuit> element to describe such circuits within the <Gate> element (see also <Circuit> below). If we think of low-level quantum hardware as programmable circuits composed of sets of universal gate, we could then imagine "compiling" a "logical" gate into its universal equivalent "hardware" version, therefore establishing the foundation of a QIS compiler.

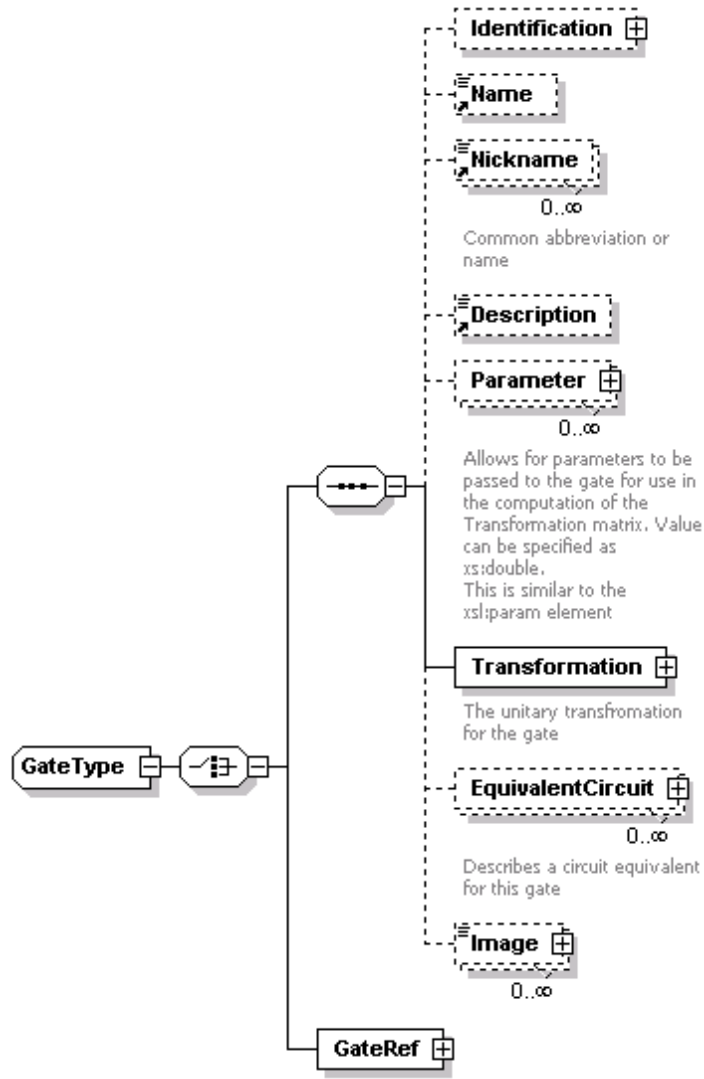

*Figure 10 - The Gate element type*

```
<Gate>
   <Identification id="Z"/>
   <Name>Pauli-Z</Name>
   <Transformation size="1">
      <Cell row="1" col="1" r="1"/>
      <Cell row="2" col="2" r="-1"/>
   </Transformation>
</Gate>

<Gate>
   <Identification id="SHIFT"/>
   <Name>Phase Shift</Name>
   <Parameter>
      <Name>theta</Name>
   </Parameter>
   <Transformation size="1">
      <Cell row="1" col="1" r="1"/>
      <Cell row="2" col="2">
         <Symbolic
syntax="html">e^(2πiθ)</Symbolic>
      </Cell>
   </Transformation>
</Gate>
```

```
<Gate>
   <Identification id="H"/>
   <Name>Hadamard</Name>
   <Transformation size="1">
      <Multiplier r="0.707106781">
         <Symbolic
syntax="odf">1/sqrt(2)</Symbolic>
         <Symbolic
syntax="html">1/sqrt(2)</Symbolic>
      </Multiplier>
      <Cell row="1" col="1" r="1"/>
      <Cell row="1" col="2" r="1"/>
      <Cell row="2" col="1" r="1"/>
      <Cell row="2" col="2" r="-1"/>
   </Transformation>
</Gate>
```

*Figure 11 - Single qubit gates examples*

```
<Gate>
   <Identification id="C-NOT"/>
   <Name>Controlled-NOT</Name>
   <Nickname>C-NOT</Nickname>
   <Transformation size="2">
      <Cell row="1" col="1" r="1"/>
      <Cell row="2" col="2" r="1"/>
      <Cell row="3" col="4" r="1"/>
      <Cell row="4" col="3" r="1"/>
   </Transformation>
</Gate>

<Gate>
   <Identification id="C-S"/>
   <Name>Controlled Phase</Name>
   <Transformation size="2">
      <Cell row="1" col="1" r="1"/>
      <Cell row="2" col="2" r="1"/>
      <Cell row="3" col="3" r="1"/>
      <Cell row="4" col="4" i="1"/>
   </Transformation>
</Gate>
```

```
<Gate>
   <Identification id="FREDKIN"/>
   <Name>Fredkin</Name>
   <Nickname>Controlled Swap</Nickname>
   <Transformation size="3">
      <Cell row="1" col="1" r="1"/>
      <Cell row="2" col="2" r="1"/>
      <Cell row="3" col="3" r="1"/>
      <Cell row="4" col="4" r="1"/>
      <Cell row="5" col="5" r="1"/>
      <Cell row="6" col="7" r="1"/>
      <Cell row="7" col="6" r="1"/>
      <Cell row="8" col="8" r="1"/>
   </Transformation>
</Gate>
```

*Figure 12 - Multiple qubit gates examples*

### 3.2.1 Parameterized Gates

As I tried to describe a "shift" gate, a special kind of <Gate> emerged as the value of one of its cell is not constant and it actually necessitates a "parameter" as an input. I'm uncertain whether this gate will ever be physically realized but its logical existence is certainly justified. To manage this case, I therefore added the ability to include one of more <Parameter> elements in the <Gate> description. Cells in the unitary transformation for which no quantitative value is provide (@r and @i attributes are missing) can then be computed using the extra parameter to evaluate the symbolic expression. This is of course to be taken into account by information systems that process this XML (remember that XML is concerned about describing objects, not processing).

```
<Gate>
    <Identification id="SHIFT"/>
    <Name>Phase Shift</Name>
    <Parameter>
        <Name>θ</Name>
    </Parameter>
    <Transformation size="1">
        <Cell row="1" col="1" r="1"/>
        <Cell row="2" col="2">
            <Symbolic syntax="html">e^(2πiθ)</Symbolic>
        </Cell>
    </Transformation>
</Gate>
```

*Figure 13 - Parameterized gate example*

### 3.2.2 Implemented Gates

Overall, the following gates have been successfully described using QIS-XML:

- Single qubit gates: Hadamard, Identity, Pauli-X, Pauli-Y, Pauli-Z, Phase, Phase Shift, Square Root of Not, $\pi/8$

- Multiple qubit gates: Controlled-NOT (2), Controlled $\pi/8$ (2), Controlled Phase (2), Controlled-Z (2), Deutsch Gate (3), Fredkin (3), Swap (2), Toffoli (3).

A few of them are illustrated below in QIS-XML format.

```
<Gate>
    <Identification id="C-NOT"/>
    <Name>Controlled-NOT</Name>
    <Nickname>C-NOT</Nickname>
    <Transformation size="2">
        <Cell row="1" col="1" r="1"/>
        <Cell row="2" col="2" r="1"/>
        <Cell row="3" col="4" r="1"/>
        <Cell row="4" col="3" r="1"/>
    </Transformation>
</Gate>
```

```
<Gate>
    <Identification id="C-S"/>
    <Name>Controlled Phase</Name>
    <Transformation size="2">
        <Cell row="1" col="1" r="1"/>
        <Cell row="2" col="2" r="1"/>
        <Cell row="3" col="3" r="1"/>
        <Cell row="4" col="4" i="1"/>
    </Transformation>
</Gate>
```

*Figure 14 – The C-NOT and Controlled Phase gates in QIS-XML*

```
<Gate>
    <Identification id="DEUTSCH"/>
    <Name>Deutsch Gate</Name>
    <Description>The Deutsch gate is a quantum gate, which is based on the idea of a
Toffoli gate. It is a 3 input gate where the two top inputs control the action of the
bottom line. But this time the action is not a toggle. Instead it is a spin rotation by
angle θ about the x axis.</Description>
    <Parameter>
        <Name>theta</Name>
    </Parameter>
    <Transformation size="3">
        <Cell row="1" col="1" r="1"/>
        <Cell row="2" col="2" r="1"/>
        <Cell row="3" col="3" r="1"/>
        <Cell row="4" col="4" r="1"/>
        <Cell row="5" col="5" r="1"/>
        <Cell row="6" col="6" r="1"/>
        <Cell row="7" col="7">
            <Symbolic syntax="html">cos(θ)</Symbolic>
        </Cell>
        <Cell row="7" col="8">
            <Symbolic syntax="html">i sin(θ)</Symbolic>
        </Cell>
        <Cell row="8" col="7">
            <Symbolic syntax="html">i sin(θ)</Symbolic>
        </Cell>
        <Cell row="8" col="8">
            <Symbolic syntax="html">cos(θ)</Symbolic>
        </Cell>
    </Transformation>
</Gate>
```

*Figure 15 – The DEUTSCH gate in QIS-XML*

## 3.3 QUANTUM CIRCUITS

With a basic <Gate> metadata structure available and a few gates described in a QIS-XML document, I then moved on to the modeling of quantum circuits. The primary objective was to be able to generate a simple visualization for circuits.

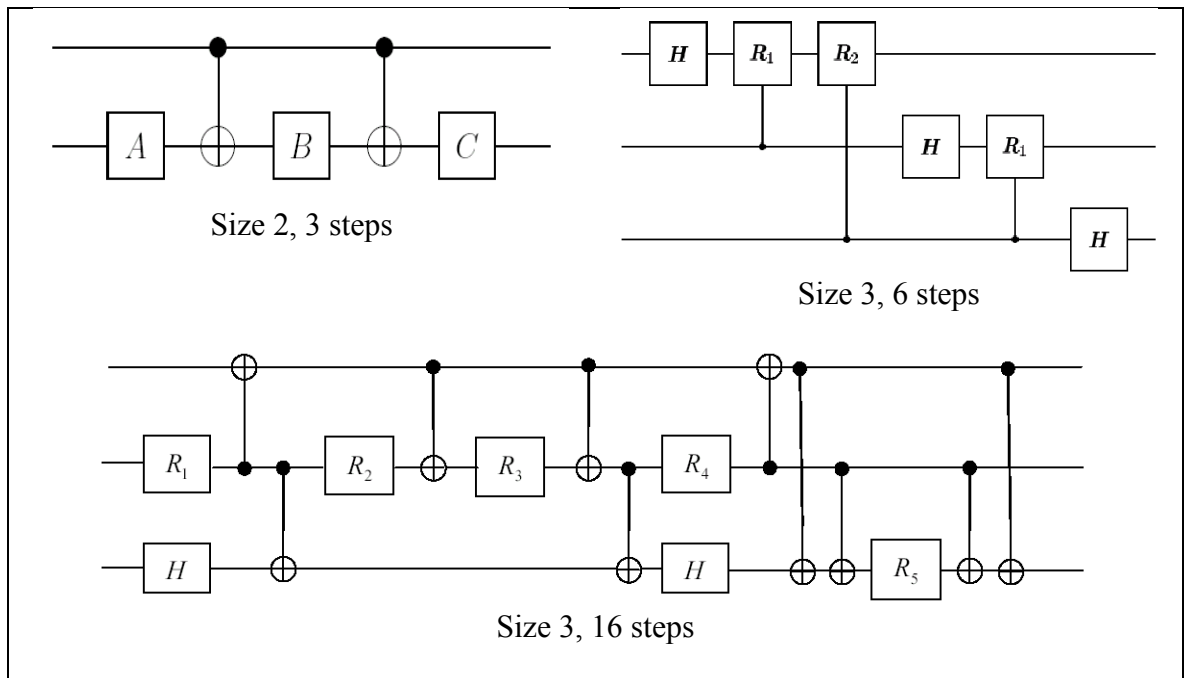

Size 2, 3 steps

Size 3, 6 steps

Size 3, 16 steps

*Figure 16 - Simple quantum circuits*

After reviewing numerous circuits from the general QIS literature, I broke down the <Circuit> concept into a collection of <Step> made of <Operation>. Steps represent the vertical cross-sections of the circuit and an operation is used to <Map> the circuit qubits to the input of a quantum <Gate> or potentially another <Circuit>. Like for gates, a few additional elements are available to provide descriptive information for the circuit

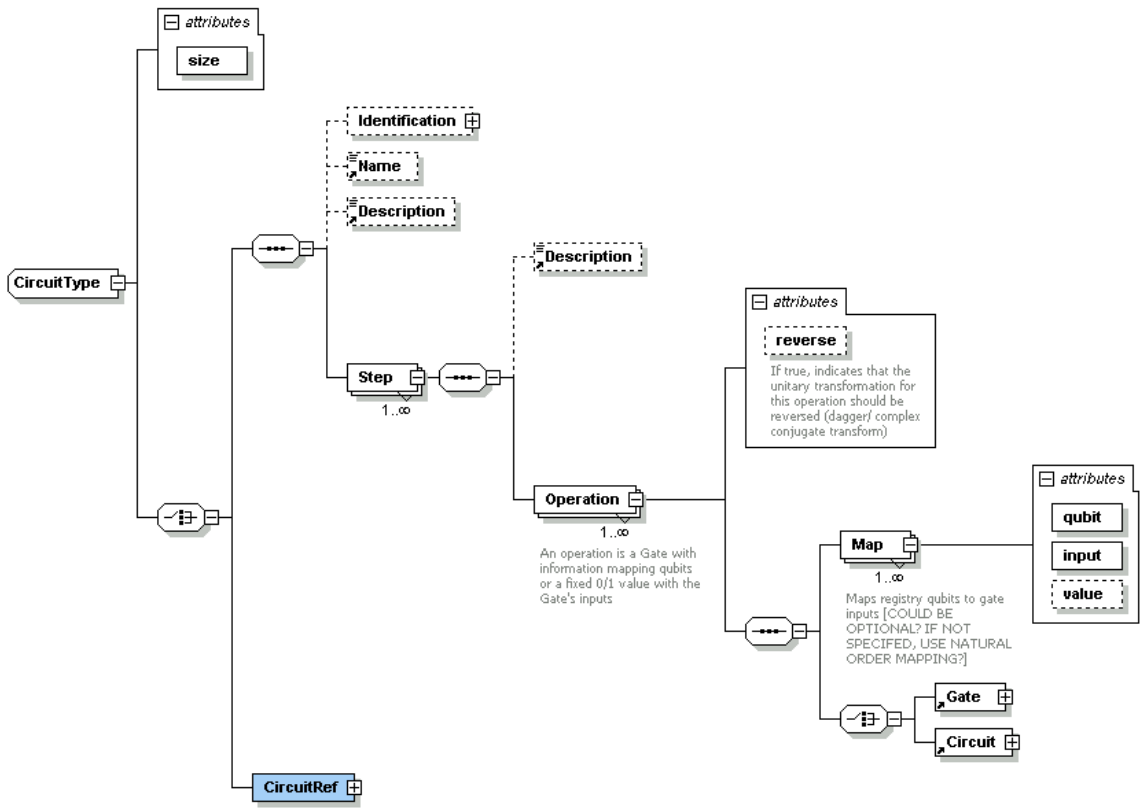

*Figure 17 - The Circuit element type*

A simple example of a circuit using the above structure is an equivalent circuit of a quantum NOT (or Pauli-X) gate composed of two square root of not gate.

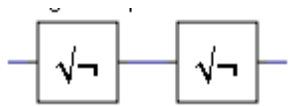

*Figure 18 - Quantum NOT equivalent circuit*

This circuit can be described in QIS-XML as follows:

```
<Circuit size="1">
    <Description>NOT gate equivalent circuit
made of two "Square root of not"
gates</Description>
    <Step>
        <Operation>
            <Map qubit="1" input="1"/>
            <Gate>
                <GateRef id="SQRT-NOT"/>
            </Gate>
        </Operation>
    </Step>
    <Step>
        <Operation>
            <Map qubit="1" input="1"/>
            <Gate>
                <GateRef id="SQRT-NOT"/>
            </Gate>
        </Operation>
    </Step>
</Circuit>
```

– The circuit size is one qubit
– It is made of two steps

– Step 1 has one operation consisting in applying a square root of not gate to the qubit. Notice the use of a GateRef element to point at the gate defined elsewhere in the XML

– Step 2 has one operation consisting in applying a square root of not gate to the qubit

*Figure 19 - Quantum NOT equivalent circuit in QIS-XML*

A slightly more complex example would be a 3-qubit phase flip circuit used for quantum error correction (Nielsen and Chuang 2000)

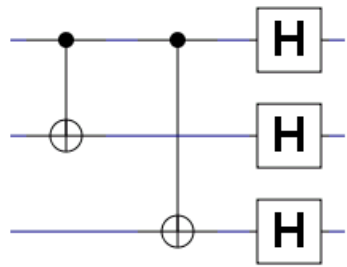

*Figure 20 - 3-qubit phase flip error correction circuit*

This circuit can be described in QIS-XML as:

```
<Circuit size="3">
   <Identification id="three_qubit_phase_flip"/>
   <Name>3-qubit phase flip code</Name>
   <Step>
      <Operation>
         <Map qubit="1" input="1"/>
         <Map qubit="2" input="2"/>
         <Gate>
            <GateRef id="C-NOT"/>
         </Gate>
      </Operation>
   </Step>
   <Step>
      <Operation>
         <Map qubit="1" input="1"/>
         <Map qubit="3" input="2"/>
         <Gate>
            <GateRef id="C-NOT"/>
         </Gate>
      </Operation>
   </Step>
   <Step>
      <Operation>
         <Map qubit="1" input="1"/>
         <Gate>
            <GateRef id="H"/>
         </Gate>
      </Operation>
      <Operation>
         <Map qubit="2" input="1"/>
         <Gate>
            <GateRef id="H"/>
         </Gate>
      </Operation>
      <Operation>
         <Map qubit="3" input="1"/>
         <Gate>
            <GateRef id="H"/>
         </Gate>
      </Operation>
   </Step>
</Circuit>
```

– The circuit is three qubit in size

– The first step consists in a single operation that applies a C-NOT gate to qubits 1 & 2

– The second step consists in a single operation that applies a C-NOT gate to qubits 1 & 3

– The third step consists in a three operations that apply an Hadamard gate to each qubit

*Figure 21 - 3-qubit phase flip circuit in QIS-XML*

A few attributes have been made available to meet special cases. For example, the <Map> element can optionally specify a @value attribute to set the input to a fixed value of zero or one. A @reverse attribute is also attached to the <Operation> element to indicate that the associated unitary transformation should actually be reversed (apply the conjugate transform). This is for example useful for the Toffoli gate equivalent circuit in which reversed pi/8 gates are used (Nielsen and Chuang 2000).

More complex examples have been described in QIS-XML such as the 9-qubit Shor code or the 3-qubit quantum Fourier transforms. The QIS-XML core library contains gate equivalent circuits like a Controlled-not gate made of two Hadamard and one Z gate or a Toffoli gate composed of Hadamard, Phase, C-Not and Pi/8.

# 4   QIS-XML VALIDATION

Having the ability to describe gates and circuits is an interesting first step but so far not very useful. It can be used by computers but, while XML is in theory human readable, this is usually not exactly what we want to show to the end users.

As mentioned in the introduction, the XML technology comes with tools to transform XML documents into other formats. This process is known as the Extensible Stylesheet Language Transformations (XSLT). It is very commonly used to convert an XML into HTML web pages but can also create PDF files, convert into other XML formats or simply output text.

Like nearly everything in the XML world, an XSLT is written using the XML language and therefore saved as an XML document, typically with the ".xsl" or ".xslt" extension. Explaining how XSLT are designed is outside the scope of this document. Numerous resources and tutorials are available on the Internet. In the context of QIS, it is interesting to note that XSL is a Turing complete language.

To create an HTML page out of an XML document (like a QIS-XML file), we basically need an XSLT that extract the relevant information from the document and describes how it fits into the HTML page layout (HTML is concerned with presentation, no content). As this basically depends on the metadata elements contained in the document, a XSL transformation is typically designed to work with a single XML specification. It however works with any document that validates against the specific schema. For example, a XSLT designed to work with QIS-XML should be able to transform any valid QIS-XML document.

In order to perform the transformation itself, we need one more thing: an application that can combine the XML source document and the XSL Transformation to produce the actual output (see illustration below). This is known as an XSL processor. XML being a global standard, such utility is actually built in your operating system and even directly available in your favorite HTML browser. Most modern programming languages (C++, Java, .NET, pHp, etc.) also have the ability to natively perform such operation.

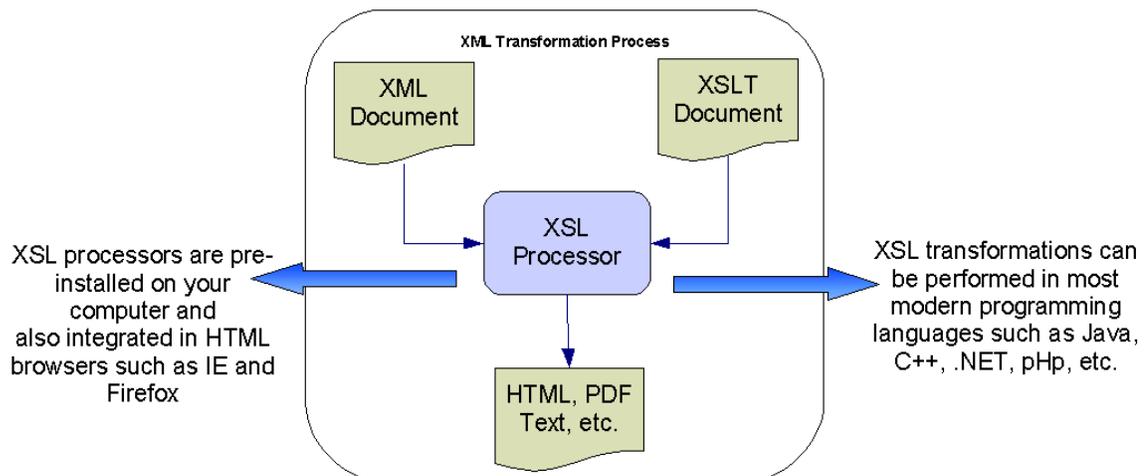

*Figure 22 - The XSL transformation process*

This means that if we have a XML document and a XSL transformation, we can basically use it on any computer without the need to install or develop software. We can also integrate it into any application we might be developing. This begins to show some of the tremendous advantages of the XML technology: It works transparently across systems and is language independent.

## 4.1   FROM QIS-XML TO HTML

With XSLT, we can now convert our QIS-XML document containing information (metadata) on gates and circuits into HTML pages. The HTML will basically be a *representation* of the XML content that can be displayed in a browser. There are of course many different "views" we can generate for our QIS-XML: list of gates, list of circuits, counting the number of gate and circuits, information on all the one qubit gates and circuits, etc. Each of these can be an individual XSLT or more complex parameterized transformations with reusable code. In the XML world, XSL is actually the closest thing to a programming language.

A useful concept to introduce here as well is that the XML document is a well-structured information container. An XSL transformation can "query" the content of the XML file to format it into an HTML for presentation to the user. This is another very powerful feature of XML: a document can be treated as a database system and therefore queried to extract information. Just like traditional databases have the Structured Query Language (SQL) to perform searches, XML is equipped with languages called XPath and XQuery that can be used to retrieve metadata from a document. The major difference is that these queries can be performed by the XSL processor and therefore you do not need to install any new software on your computer or even design the database! This functionality is build in the technology and freely available.

The examples provided in this document have all been built using XSLT. A simple one illustrated below shows gates in the document along with a representation of their unitary transformation matrix (the XSL code itself is too long to include here).

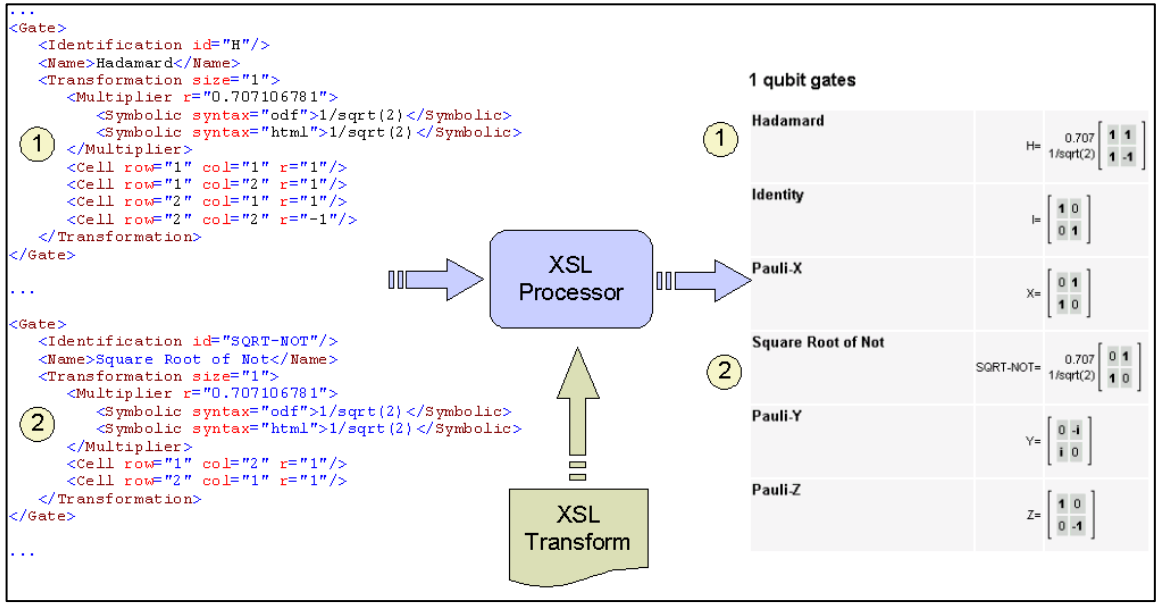

*Figure 23 - Transformation of an XML document into HTML using XSLT and an XSL Processor*

### Deutsch Gate

The Deutsch gate is a quantum gate, which is based on the idea of a Toffoli gate. It is a 3 input gate where the two top inputs control the action of the bottom line. But this time the action is not a toggle. Instead it is a spin rotation by angle θ about the x axis. See also http://beige.ucs.indiana.edu/M743/node95.html.

DEUTSCH(theta)=

| 1 | 0 | 0 | 0 | 0 | 0 | 0 | 0 |
|---|---|---|---|---|---|---|---|
| 0 | 1 | 0 | 0 | 0 | 0 | 0 | 0 |
| 0 | 0 | 1 | 0 | 0 | 0 | 0 | 0 |
| 0 | 0 | 0 | 1 | 0 | 0 | 0 | 0 |
| 0 | 0 | 0 | 0 | 1 | 0 | 0 | 0 |
| 0 | 0 | 0 | 0 | 0 | 1 | 0 | 0 |
| 0 | 0 | 0 | 0 | 0 | 0 | cos(θ) | i sin(θ) |
| 0 | 0 | 0 | 0 | 0 | 0 | i sin(θ) | cos(θ) |

*Figure 24 - HTML representation of the Deutsch Gate*

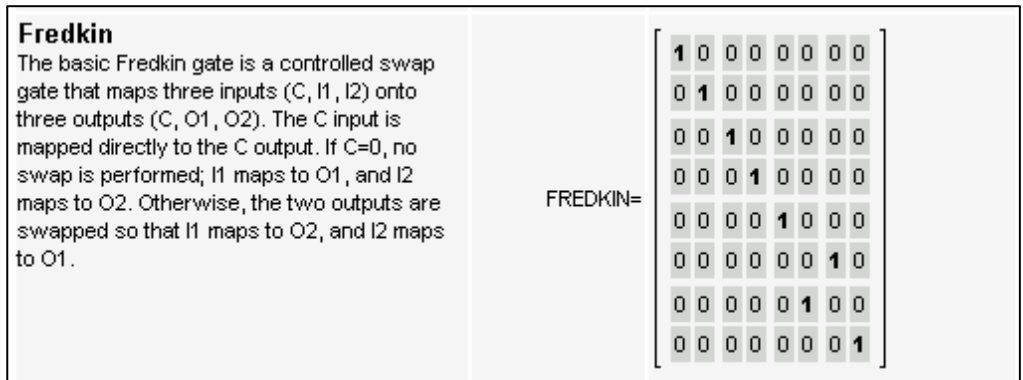

*Figure 25 - HTML representation of the Fredkin Gate*

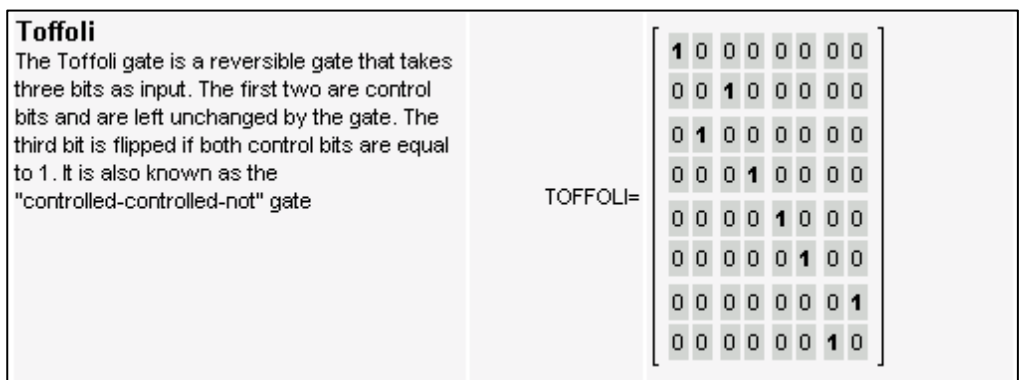

*Figure 26 - HTML representation of the Toffoli Gate*

## 4.2   QIS-XML VALIDATION

Transformation can also be used for other purposes such as reporting and validation. Fundamentally, performing validation consists in parsing the XML content to check its consistency and coherence and produce a report out of it. If the report happens to be in HTML format, it can be viewed in a browser. It is in essence the same as a transformation of the XML metadata.

Some of the basic document validation can actually be taken care of by the XML schema itself. For example, in the model, we can specify that a <Gate> *must* contain a single <Transformation> element or can enforce that the @input attribute of a <Map> element must be a positive integer. When we validate a QIS-XML document against the QIS-XML schema (see the introduction), XML will tell us if something is fundamentally wrong in my document. The XML Schema language, however, has a limitation and cannot check the integrity of quantum gates and circuits.

To implement more complex validation procedures, I therefore created a transformation that parses the QIS-XML document and looks for errors like:

- Qubit whose total probability is not equal to 1

- Unitary transformation in gates with cells outside the matrix row/col range

- Qubit value range in circuit operation (cannot be greater than the circuit size)

- Duplicate qubit or input mappings in circuit operations and steps

- etc.

The output of the validation transformation is an HTML document that reports errors and warnings for each <Gate> and <Circuit>. An example is shown below for the 9-qubit Shor code circuit (in which an error has been introduced). I expect the validation transform to grow as the QIS-XML increases in complexity.

```
Circuit id0x0606e4d8, Size 9, 5 step(s)
9-qubit Shor quibit code
Encoding circuit for the Shor nine qubit code.
  Step 1, 1 operation(s)
    Warning: Not all qubits have been mapped.
    1: Controlled-NOT (C-NOT) [1=1,4=3]
        ERROR: Map 1 input=3 is out of Gate range.
  Step 2, 1 operation(s)
    Warning: Not all qubits have been mapped.
    1: Controlled-NOT (C-NOT) [1=1,7=2]
  Step 3, 3 operation(s)
    Warning: Not all qubits have been mapped.
    1: Hadamard (H) [1=1]
    2: Hadamard (H) [4=1]
    3: Hadamard (H) [7=1]
  Step 4, 3 operation(s)
    Warning: Not all qubits have been mapped.
    1: Controlled-NOT (C-NOT) [1=1,2=2]
    2: Controlled-NOT (C-NOT) [4=1,5=2]
    3: Controlled-NOT (C-NOT) [7=1,8=2]
  Step 5, 3 operation(s)
    Warning: Not all qubits have been mapped.
    1: Controlled-NOT (C-NOT) [1=1,3=2]
    2: Controlled-NOT (C-NOT) [4=1,6=2]
    3: Controlled-NOT (C-NOT) [7=1,9=2]
```

*Figure 27 – Example of a validation output report*

## 4.3   INTERIM RESULTS

With the QIS-XML schema and a few XSL transformations at hand, we therefore now
have the ability to describe most quantum gates and circuits in a standard format and
display their characteristics in HTML or user friendly representations. The information
can also be queried through any dimension using the XPath language to various extract
elements of metadata. Such information can be accessed by any operation system and be
understood by most modern programming languages. As we will briefly discuss later, we
also have the ability to easily exchange it across network such as the Internet or publish it
in metadata registries

# 5 VISUALIZATION USING SCALABLE VECTOR GRAPHICS (SVG)

One of the main objectives for this project was to be able to provide a graphical representation of gates and circuits. Again, when we think about it, this is conceptually a transformation of the QIS-XML document into a diagram or an image. The natural XML reflex is therefore to think XSLT but the problem in this case is that image file are often proprietary binary formats (GIF, JPG, PNG, TIF, etc.). While XSLT could in theory produce such output, a specialized program developed using Java or C would be more efficient (and more appropriate).

Fortunately, XML comes again to the rescue. There is a W3C XML specification called Scalable Vector Graphics (SVG) that can be used to describe two-dimensional vector graphics. When you think of a quantum circuit diagram, it can certainly be broken down into a collection of lines, circle, boxes and text. We should therefore be able to "describe" the diagram as a composition of these basic elements, and describing things is what XML does well.

Since for the purpose of this project I wanted to use solely XML technologies, I settled for an SVG-XML based solution for the representation of circuits.

## 5.1  WHAT IS SVG?

Numerous resources are available on the Internet on SVG-XML. While it provides a set of advanced functionalities, basic SVG is fairly straightforward to use. Simple examples are illustrated below.

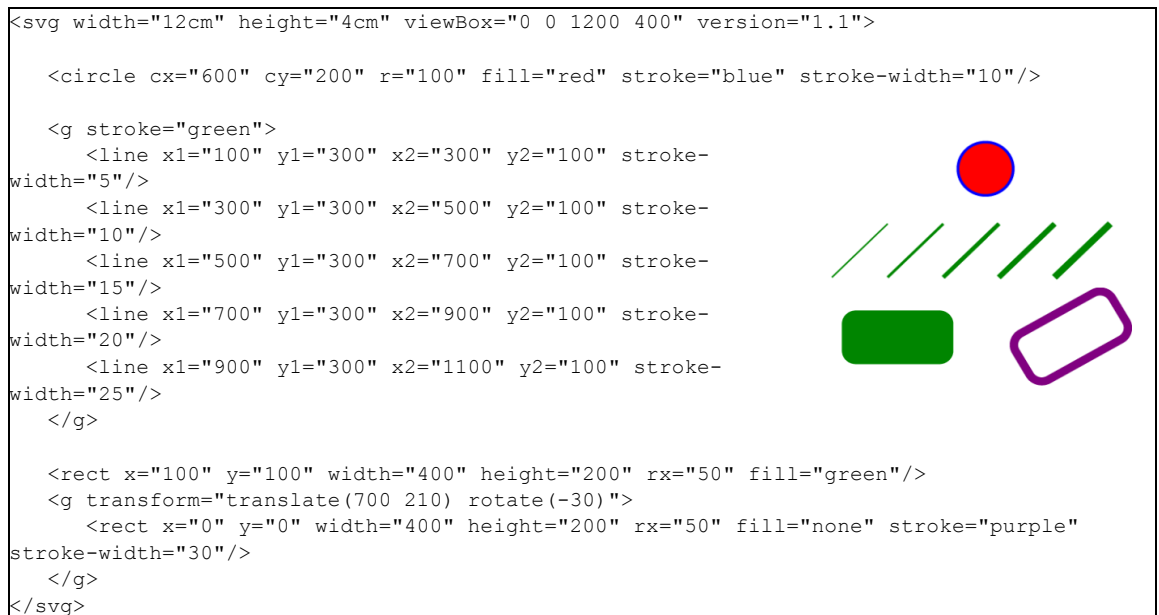

```
<svg width="12cm" height="4cm" viewBox="0 0 1200 400" version="1.1">

   <circle cx="600" cy="200" r="100" fill="red" stroke="blue" stroke-width="10"/>

   <g stroke="green">
      <line x1="100" y1="300" x2="300" y2="100" stroke-
width="5"/>
      <line x1="300" y1="300" x2="500" y2="100" stroke-
width="10"/>
      <line x1="500" y1="300" x2="700" y2="100" stroke-
width="15"/>
      <line x1="700" y1="300" x2="900" y2="100" stroke-
width="20"/>
      <line x1="900" y1="300" x2="1100" y2="100" stroke-
width="25"/>
   </g>

   <rect x="100" y="100" width="400" height="200" rx="50" fill="green"/>
   <g transform="translate(700 210) rotate(-30)">
      <rect x="0" y="0" width="400" height="200" rx="50" fill="none" stroke="purple"
stroke-width="30"/>
   </g>
</svg>
```

*Figure 28 – Basic SVG examples*

Note that while most web browsers provide some level of support for SVG, the level of implementation and compliance varies. In some case (like for Internet Explorer prior to version 9), the ability to visualize SVG may require in the installation of a plugin (the Adobe SVG viewer being a popular one).

## 5.2 QUANTUM GATES AND CIRCUITS IN SVG

Converting the QIS-XML into SVG-XML was not particularly a straightforward process. The representation of single qubit gates was simple but the automatic layout of a circuit diagram was much more problematic. Issues such as connecting lines between gate's inputs/outputs, moving multi-qubit gate's nodes to the proper location, or fixed map @input value, required significant development. The implementation currently achieved works for circuits using well-defined gates but will need improvements in order to manage more complex cases (such as circuits in circuit). As mentioned before, the XSLT language also its limitations. For complex cases, a more flexible programming language would be more appropriate. Nevertheless, the project objective of simple circuits representation was achieved. Sample outputs are shown below:

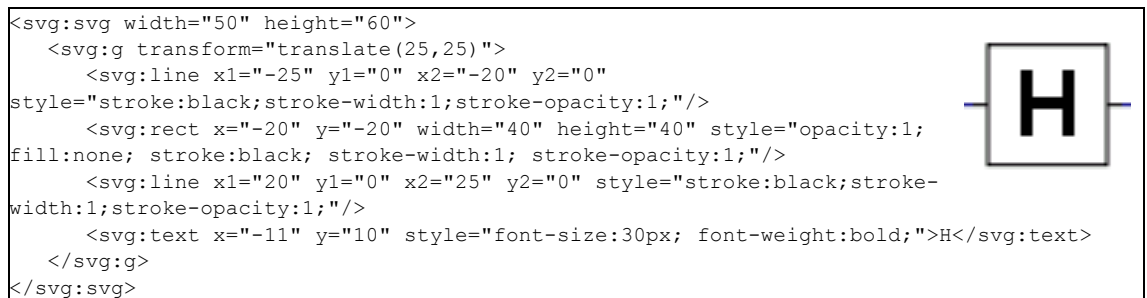

*Figure 29 - SVG-XML representation of a Hadamard Gate*

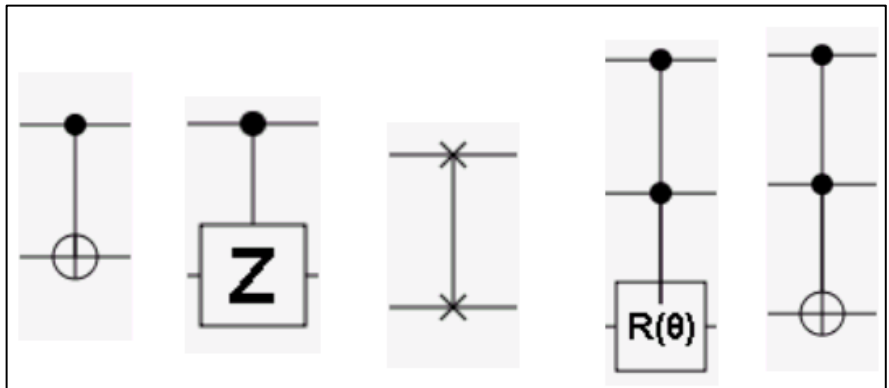

*Figure 30 - SVG-XML rendering of common QIS-XML quantum gates*

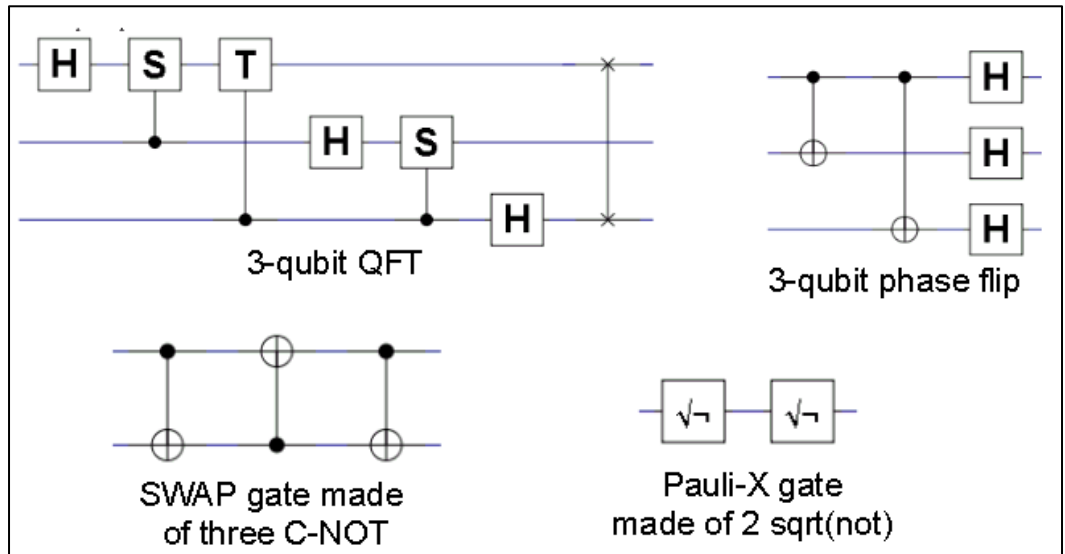

3-qubit QFT

3-qubit phase flip

SWAP gate made
of three C-NOT

Pauli-X gate
made of 2 sqrt(not)

*Figure 31 – SVG-XML rendering of simple QIS-XML circuits*

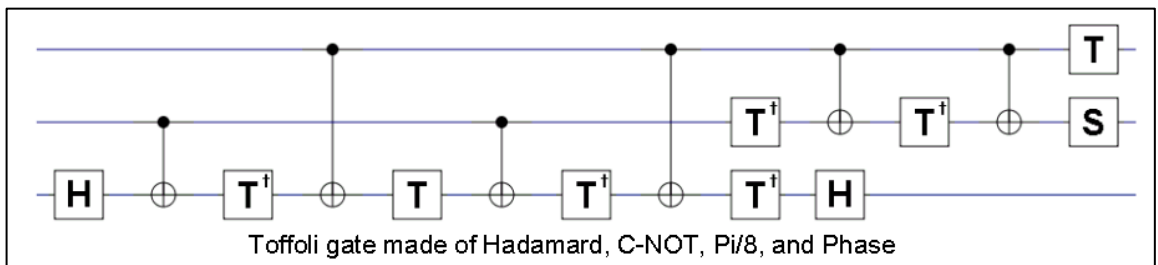

Toffoli gate made of Hadamard, C-NOT, Pi/8, and Phase

*Figure 32 –SVG-XML representation of a QIS-XML Toffoli gate equivalent circuit*

# 6   QIS-XML ENHANCEMENTS

## 6.1   THE MODULAR APPROACH

Up to this point, QIS-XML consists of a single specification used to describe gates and circuits. While it provides a good starting point, it was not initially designed to operate in a distributed / federated environment where different organizations or individuals maintain different components of the metadata. This has become a fairly common situation for XML specifications and it can be resolved through modularization.

The reasons we need this for QIS-XML are multiple:

-   The set of core quantum gates is fairly well defined and there is little need to redefine them all the time.

-   Many circuits will be designed by different organizations or individuals and should be remotely accessible and reusable

-   Quantum programs should be able to reference quantum circuits described anywhere, (not only locally)

-   The metadata should be able to reside at multiple locations and put together on demand when the need arises.

We therefore introduce at this point a reorganization of QIS-XML into a more modular specification fit for use in a federated environment.

## 6.2 MODULES

Modularization of the specification consist in two steps:

- Breaking it down is multiple specifications (one per module)

- Implement a referencing mechanism

QIS-XML is therefore reorganized as follows:

- A Gate Library module to describe the core quantum gates

- A Circuit Library module used to describe one or more quantum circuits

- A new Program Library module to capture metadata for programming. This module is further described below but basically captures metadata on quantum registers and circuits

- A Reusable Components module that defines metadata elements common to the above modules

- A top level Instance module necessary to bring together Gate, Circuit and Program libraries

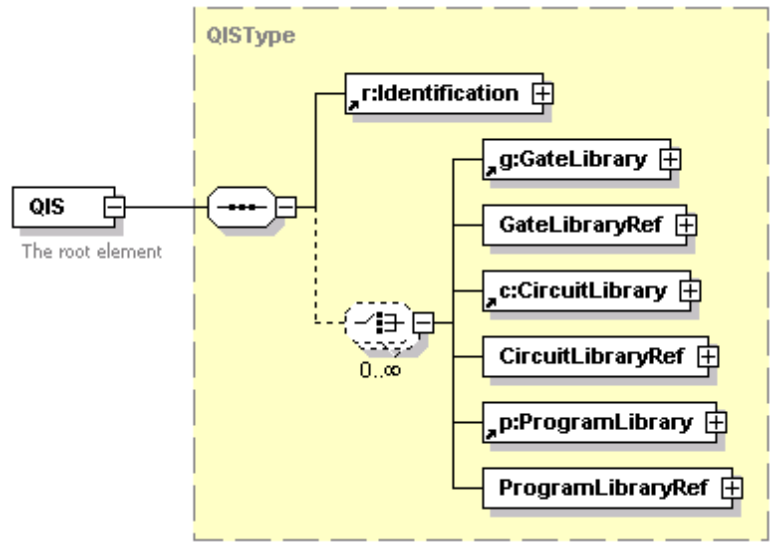

*Figure 33 – Modularized version of QIS-XML*

## 6.3  NAMESPACES

### 6.3.1   Namespace identifiers

In order to uniquely identify elements belonging to a specific schema (module), XML uses what are called "namespaces". This ensures that elements with the same name can be properly referenced and co-exists. For the enhanced version of QIS-XML, the following namespaces have been defined:

| Module | Namespace | Prefix |
|--------|-----------|--------|
| Instance | qis:instance:1_0 | i |
| Gate | qis:gate:1_0 | g |
| Circuit | qis:circuit:1_0 | c |
| Program | qis:program:1_0 | p |
| Reusable | qis:reusable:1_0 | r |

*Figure 34 – QIS-XML namespaces*

### 6.3.2   The referencing mechanism

In the initial version of QIS-XML, identifiable object were uniquely determined using a standard xs:ID attribute. The modularization of the specification and the fact the elements can make references to external entities now require more granularity. An advanced referencing mechanism is therefore introduced, composed of two reusable complex types:

- An <Identification> type to uniquely identify an entity and specify the agency (organization, individual) it belongs to as well as maintain multiple versions.

- A <Reference> type to point to an entity associated with the above <Identification> element. An optional @URI attribute is used for the case when the information is stored in an external document. The LibraryID, AgencyID and Version element are likewise optional and used only when necessary.

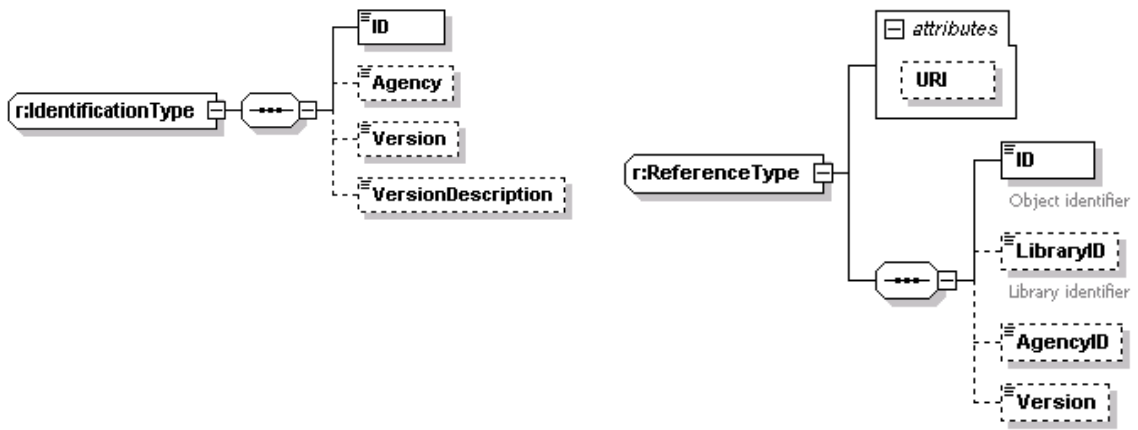

*Figure 35 – The identification and reference element types*

## 6.4 THE IMPROVED GATE AND CIRCUIT MODULES

These two modules are basically quite similar to the Gate and Circuit types that had been defined in the prototype version of QIS-XML. They are used to describe quantum gates and to design circuits based on these gates. The main difference is that they now have their own namespace and can be defined independently. This means that different agencies can design quantum circuits based on the standard set of quantum gates (no need to redefine) and that these do not need to reside in the same XML document.

To allow for multiple gates and circuits to be grouped in the same document, the <GateLibrary> and <CircuitLibrary> types have been defined as root elements.

One important modification made to the <Gate> type is that the equivalent circuit element is no longer a child of the gate but rather needs to be defined in a separate <CircuitLibrary> instance.

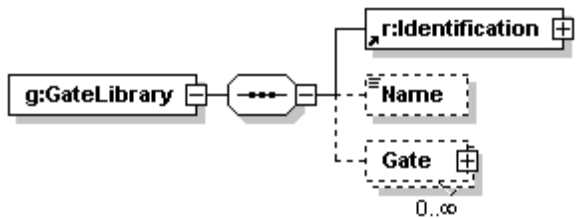

*Figure 36 – The GateLibrary element type*

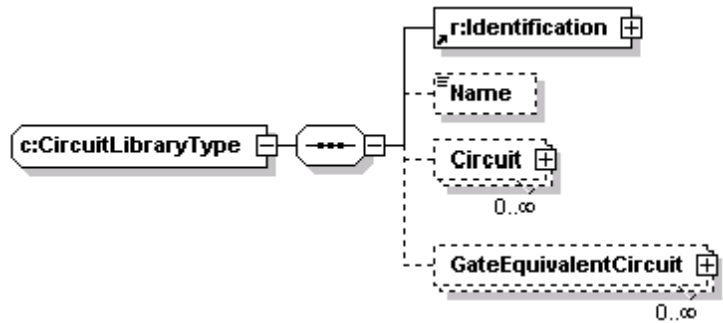

*Figure 37 – The CircuitLibrary element type*

## 6.5 THE PROGRAM MODULE

One of the initial objectives of QIS-XML was to be able to describe basic quantum algorithms in XML as a hardware/software independent framework for the implementation of quantum compilers.

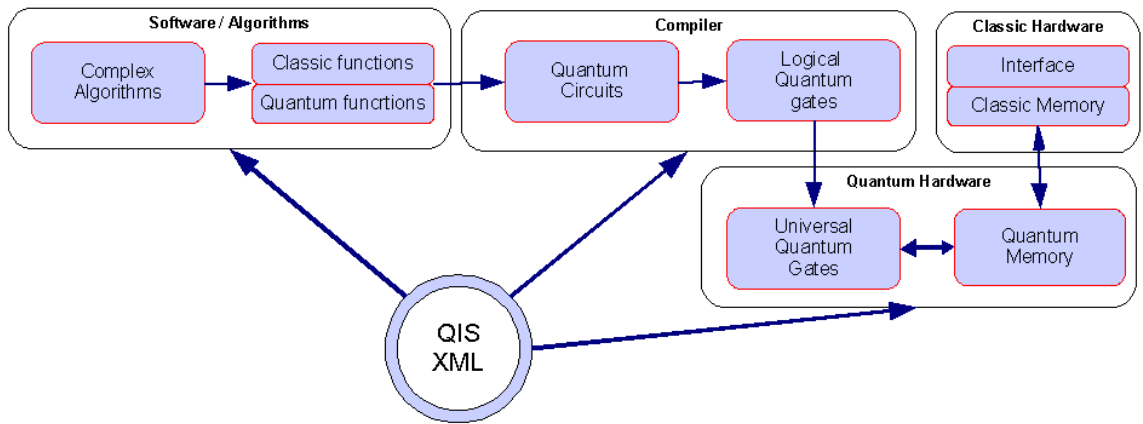

*Figure 38 – QIS-XML Compilation Model*

The general idea is that quantum hardware will likely be based on a set of universal gates acting on quantum memory. Programs however should not depend on specific sets of gates but rather on the quantum gates typically described in the QIS literature. These Logical gates are interconnected to form quantum circuits that in turn can be combined to implement algorithm. A program compilation process would then consist in converting the program into circuits, then logical gates, and then hardware specific set of universal gates that act on quantum memory.

We therefore introduce a <Program> module to support such feature. Like for gates and circuits, programs can be grouped in a program library.

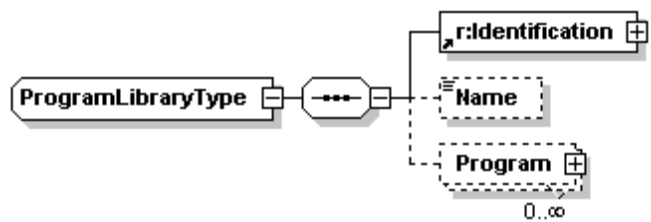

*Figure 39 – The ProgramLibrary element type*

### 6.5.1   The Program

A program brings together two main entities: the algorithm to be executed and the memory it acts upon.  The memory is a collection of qubits that can be organized in various registers. The algorithm is a collection of circuit executions and measurements (with the simple case being a single circuit and a single measurement)

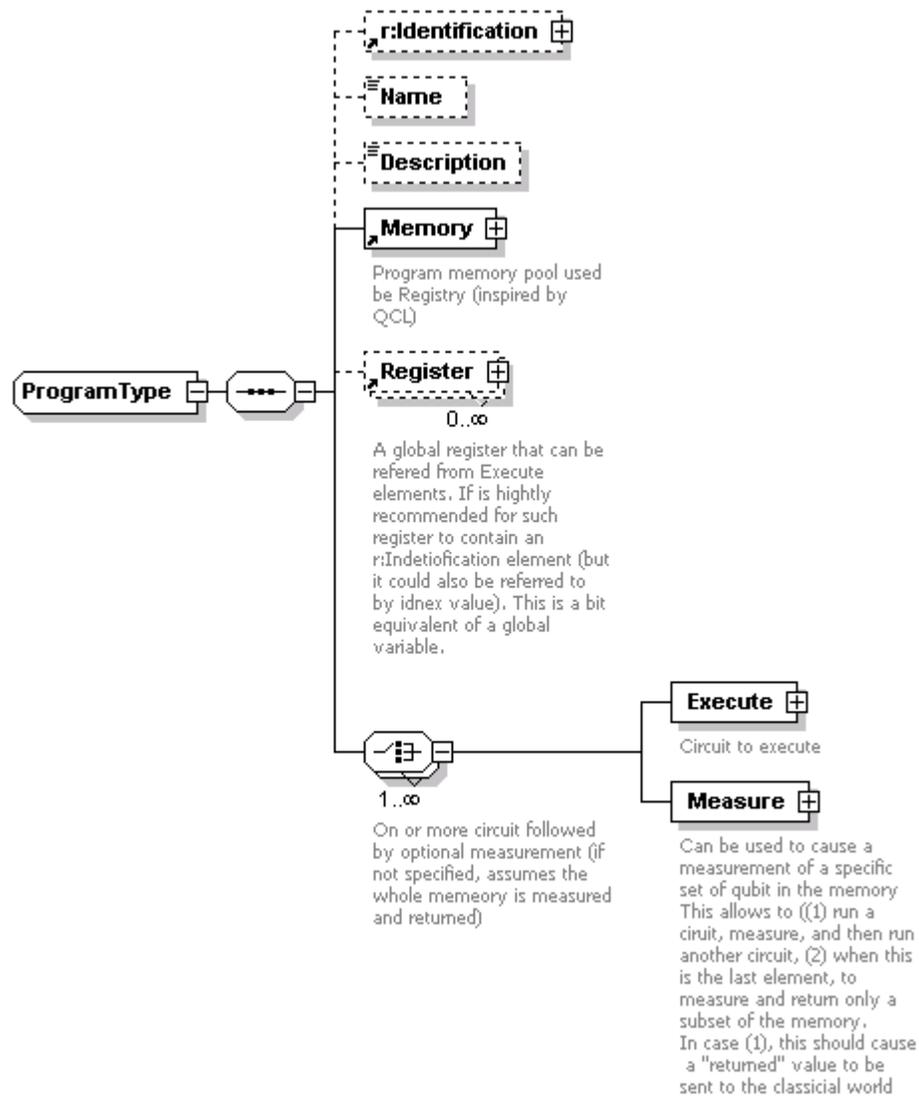

*Figure 40 – The Program element type*

### 6.5.2 Memory

I Initially looked at a program memory as a simple quantum register but was later influenced by Bernhard Omer QCL programming language (Omer, Structured Quantum Programming 2009) whose memory management is very flexible. In QCL, a global quantum memory of N qubits is made available and multiple quantum registers acting on a subset of the available qubit are defined (which somewhat provides an equivalent of classic "variables"). A similar model was also presented in (Bettelli, Serafini and Calarco 2001). This approach has therefore been taken in QIS-XML where we define a <MemoryType> and a <RegisterType>.

The only required element for Memory is its "size" attribute in number of qubits. Like many elements in QIS-XML, it has an <Identification> element and a <Name>. A <MemoryType> also comes with a <Prepare> element (further described under <Register> below) and a collection of <Qubit> elements that can actually be used to stored values if needed. Note that the <Qubit> element here is an extended version of the reusable <r:Qubit> type that also includes an @index attribute to capture the memory location (1-based index).

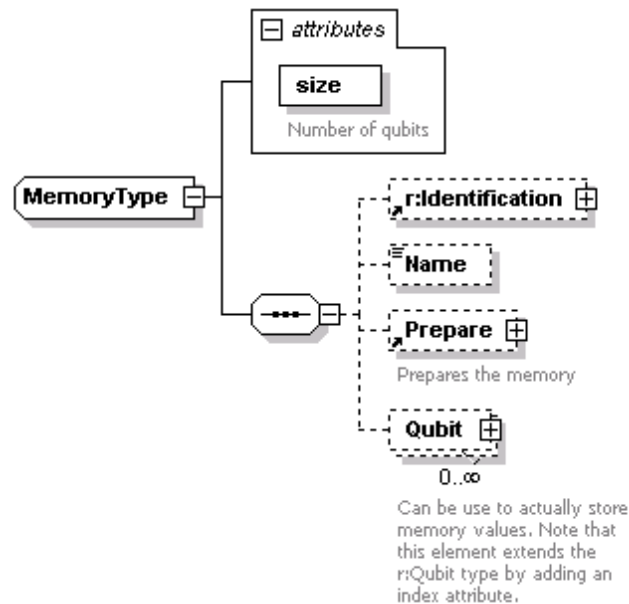

*Figure 41 – The Memory element type*

### 6.5.3 Register

A <Register> is conceptually similar to a variable and can be defined globally (as a child of the program element) or within an <Execute> element (where it is mandatory). A <Register> is simply a subset of the program <Memory>. It has a mandatory @size attribute in qubits and a set of elements used to describe which qubits of the <Memory> it refers to.

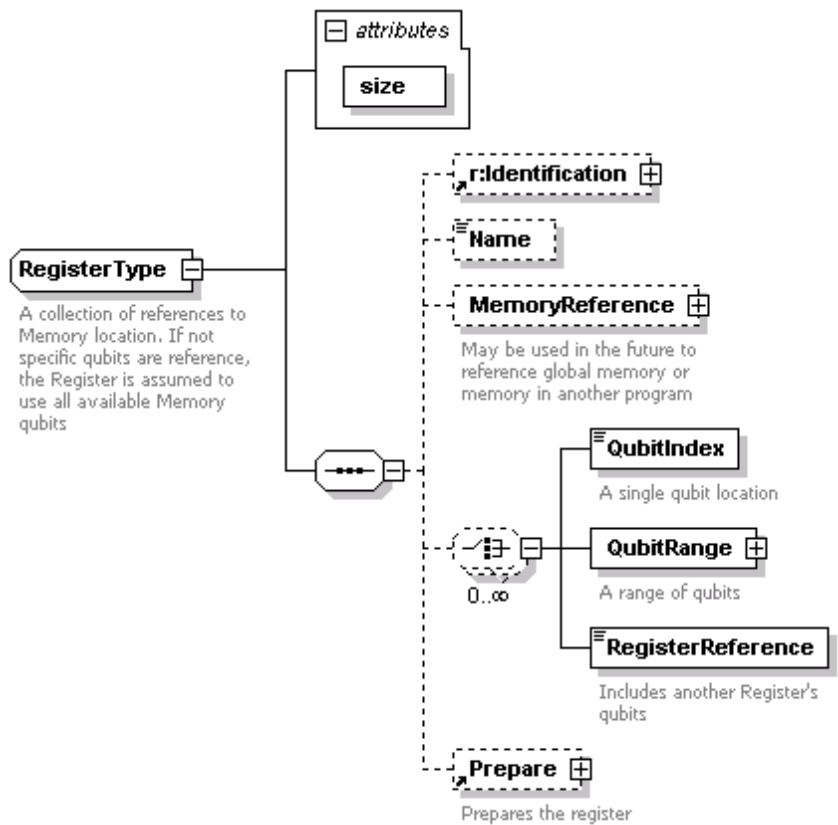

*Figure 42 – The Register element type*

There are three ways a register can make references to section of the main program memory:

- The <QubitIndex> element can be used to refer to a specific qubit

- The <QubitRange> element can be used to refer to a range of qubits using its <StartQubit> and <EndQubit> children elements

- The <RegisterReference> can be used to simply refer to another Register definition

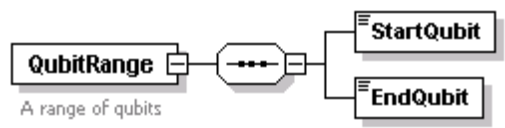

*Figure 43 – The QubitRange element type*

Note that combinations of the above three elements are allowed and that all index references in QIS-XML are 1-based (the first qubit is qubit 1, not 0). QIS-XML also makes the assumption that if no index or range is specified, it addresses the whole Memory starting from qubit 1 up to the @size of the register.

To set values of qubits in the register, we use the <Prepare> element.

<Prepare> is a collection of one of more <QubitSet> statements that can be used to set the <Value> of specific qubit of the Register. It uses the same referential mechanism as for <Memory> except that here the index values are in relation to the <Register> (not the memory).

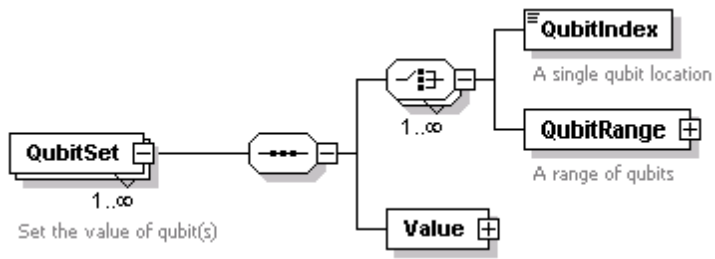

*Figure 44 – The QubitSet element type*

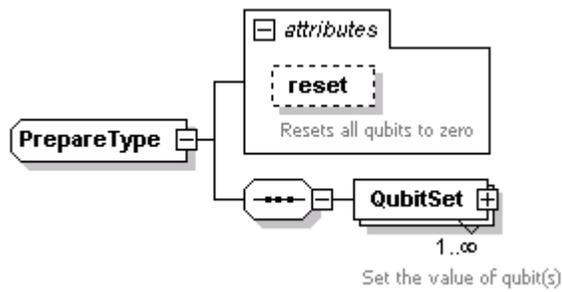

*Figure 45 – The Prepare element type*

The following register references the first 6 qubits of the main memory and sets qubits 1 and 2 to the real value 1.

```xml
<p:Register size="6">
    <p:Prepare>
        <p:QubitSet>
            <p:QubitIndex>1</p:QubitIndex>
            <p:QubitIndex>2</p:QubitIndex>
            <p:Value r="1"/>
        </p:QubitSet>
    </p:Prepare>
</p:Register>
```

*Figure 46 – Example of a Register initialization*

### 6.5.4   Execute

The <Execute> element is where the memory and circuits come together. Its basic version of composed of a mandatory <Register> element and a <Circuit>. The register, who must have the same size of the circuit, points to the qubits in memory that the circuit is acting upon. The circuit can be described directly as a child element or simply be a reference to a existing circuit defined elsewhere in a circuit library.

An experimental <Program> (or <ProgramRef>) element as also been included as, in theory, an Execute step could call a subprogram. Memory management can however become tricky in this case and this option will need to be further explored.

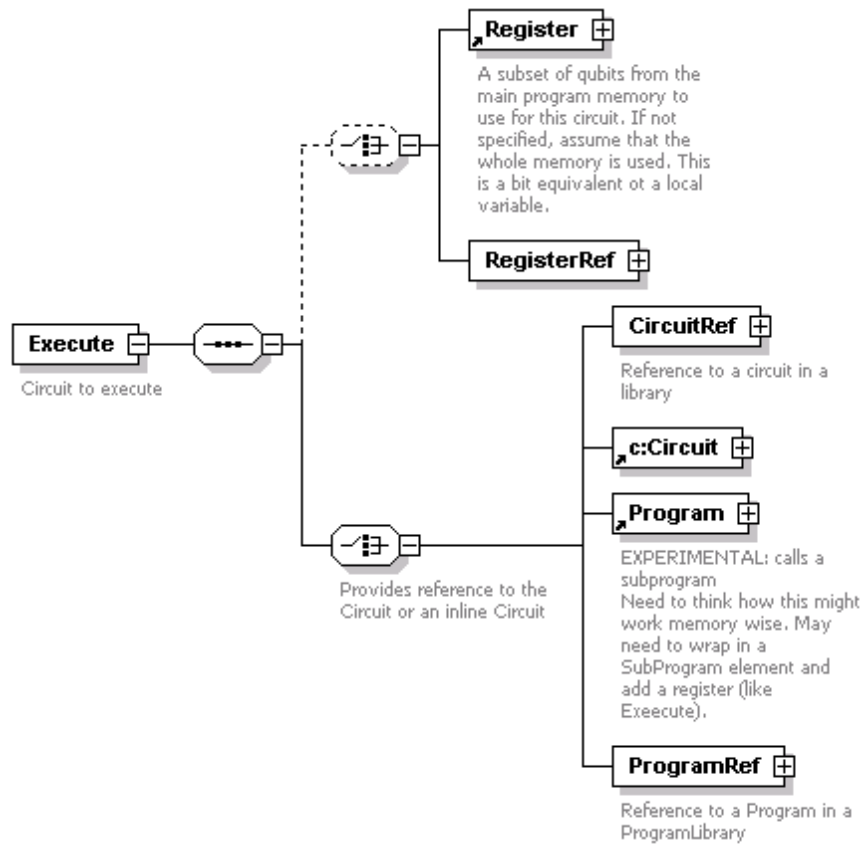

*Figure 47 – The Execute element*

### 6.5.5 Measure

The <Measure> element in a program (not to be confused in <Measurement> under <Circuit>) can be used to specify qubit measurements after of between <Execute> steps. By default, if no <Measure> element is specified, QIS-XML assumes that all qubits are measured and returned at the end of the <Program> execution.

Measurement can be used in two cases:

- Specify which qubit in memory should be measure at the end of the program. This is desirable to reduce the number of required measurements or to only measure the relevant qubits (error correction or ancillary qubits can be ignore and may not be of interest)

- If a program is composed of multiple <Execute> steps, it might be desirable to perform measurements between the Execute elements.

### 6.5.6 Program Example

The example below is a simple program that performs a 2+1 operation using a 2-qubit adder circuit (defined elsewhere in a library). This requires a 6-qubit <Memory> and contains a single <Execute> step. The 2-qubit circuit qubit 1 and 4 are used to specify the first number and qubit 2 and 5 the second. We therefore set qubit 4 and qubit 2 to specify the decimal number 2 and 1 respectively. Details on this particular adder circuit are presented as a use case later in this document.

```
<!-- 2 qubit adder 2+1 -->
<p:Program>
    <r:Identification>
        <r:ID>two_plus_one</r:ID>
    </r:Identification>
    <p:Name>Two plus One</p:Name>
    <p:Memory size="6"/>
    <p:Execute>
        <p:Register size="6">
            <p:Prepare>
                <p:QubitSet>
                    <p:QubitIndex>2</p:QubitIndex>
                    <p:QubitIndex>4</p:QubitIndex>
                    <p:Value r="1"/>
                </p:QubitSet>
            </p:Prepare>
        </p:Register>
        <p:CircuitRef>
            <r:ID>adder2</r:ID>
        </p:CircuitRef>
    </p:Execute>
</p:Program>
```

*Figure 48 - A 2-qubit adder program in QIS-XML*

## 6.6 ENHANCEMENTS SUMMARY

The enhanced version of QIS-XML has been upgraded to a more useful specification that allows for distributed or federated maintenance of quantum gates, circuits and pseudo-code metadata.

The addition of a <Program> module to QIS-XML provides a flexible way to describe quantum algorithm based on circuits and gates. The advanced capabilities of the module remain to be explored. We need however to be cautious in not making the specification too flexible as it may not be possible for quantum hardware or simulator to support such advanced features. I however at this time want to leave the door open rather than restrict functionality based on today's constraints.

# 7 A CASE EXAMPLE OF QIS-XML

With the enhanced version of QIS-XML, we can now demonstrate through a simple case how XML can be at the foundation of a quantum programming architecture.

The example presented here consists in creating a quantum adder circuit in QIS-XML by using a modified version of GenAdder (a small utility software designed by NIST in 2003 to generate quantum circuits) and, through basic XSL transformations, compile the adder program into code that can be executed on a quantum computer simulator. The visualization transforms developed earlier are also used to produce the circuit diagrams in SVG format.

The purpose is to illustrate through a simple example that circuits and programs can be described in QIS-XML and transparently executed on different quantum computing platforms. We use simulators at this time as no quantum hardware is readily available. Adding such new platform would however simply consist in preparing a new "compiler" XSL transformation to convert the XML into the proprietary scripting language.

## 7.1 NIST GENADDER

### 7.1.1 Overview

GenAdder is a simple quantum circuit generator developed in 2003 by Dr. Paul E. Black at NIST (note that this application is no longer available on the Internet). It creates quantum circuit to add two binary numbers of any width using Toffoli and CNOT gates. The circuit generated is a ripple-carry adder. It follows the block scheme described by Jozef Gruska (Gruska 2000) and was optimized by Paul E. Black. For N bit wide arithmetic, the circuit uses 3N qubits: 2N for inputs, N-1 ancilla, and a carry out. One of the inputs serves as the output sums. It is assumed that there is no carry in. Although it uses ancil (Gruska 2000)la, they are reset to 0, thus there is no "garbage". The circuit uses 7N-6 gates, N > 1.

A simple circuit example is a 2-bit wide adder is shown below. It uses six qubits: inp0, inp1, sum0, sum1, anc0, and carryOut. The input is inp0/1 and sum0/1. The output is sum0, sum1, and carryOut. The least significant bits are at the top.

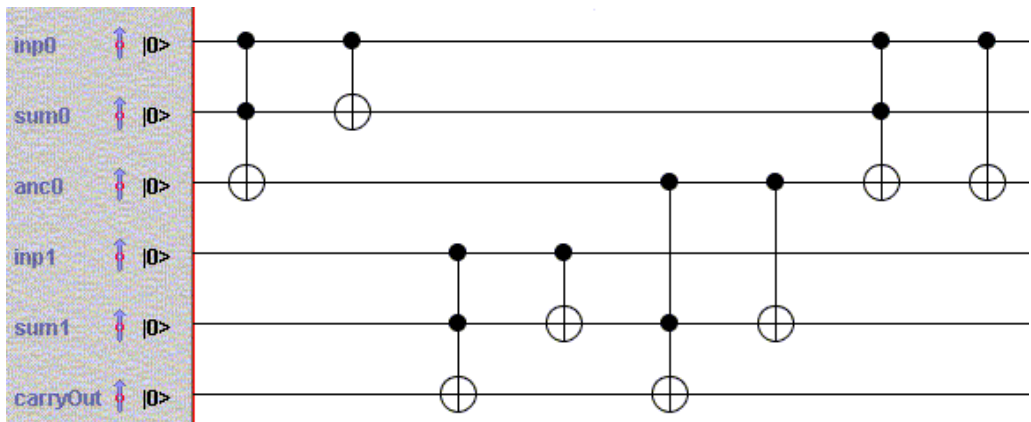

*Figure 49 – A 2-bit wide adder using GenAdder*

GenAdder was written in C and outputs a proprietary text description of the circuit as illustrated below. The program takes a single parameter as input to specific the number of desired qubits.

```
qubit_0="inp0"
qubit_1="sum0"
qubit_2="anc0"
qubit_3="inp1"
qubit_4="sum1"
qubit_5="anc1"
qubit_6="inp2"
qubit_7="sum2"
qubit_8="carryOut"
gate0={1:1:NOT:-:-:-:-:-:-}
gate1={1:NOT:-:-:-:-:-:-:-}
gate2={-:-:-:1:1:NOT:-:-:-}
gate3={-:-:-:1:NOT:-:-:-:-}
gate4={-:-:1:-:1:NOT:-:-:-}
gate5={-:-:-:-:-:-:1:1:NOT}
gate6={-:-:-:-:-:-:1:NOT:-}
gate7={-:-:-:-:-:1:-:1:NOT}
gate8={-:-:-:-:-:1:-:NOT:-}
gate9={-:-:1:-:1:NOT:-:-:-}
gate10={-:-:-:1:1:NOT:-:-:-}
gate11={-:-:-:1:-:NOT:-:-:-}
gate12={-:-:1:-:NOT:-:-:-:-}
gate13={1:1:NOT:-:-:-:-:-:-}
gate14={1:-:NOT:-:-:-:-:-:-}

# Quantum circuit to add 3 qubits without garbage
# Produced by genadder version 1.0
# $Header: /home/black/Quantum/GenAdder/RCS/genadder.C,v 1.1 2003/08/20
13:28:20 $
#   For more information, contact Paul E. Black
# Command line:
#   3

# Declare qubits
variable inp0, sum0, anc0, inp1, sum1, anc1, inp2, sum2, carryOut: qubit;
= |000000000>;

#   The adder itself
# sum bit 0
c2not(inp0, sum0, anc0);
cnot(inp0, sum0);
# sum bit 1
c2not(inp1, sum1, anc1);
cnot(inp1, sum1);
c2not(anc0, sum1, anc1);
# sum bit 2
c2not(inp2, sum2, carryOut);
```

```
cnot(inp2, sum2);
c2not(anc1, sum2, carryOut);
# finish bit 2
cnot(anc1, sum2);
# finish bit 1
c2not(anc0, sum1, anc1);
c2not(inp1, sum1, anc1);
cnot(inp1, anc1);
cnot(anc0, sum1);
# finish bit 0
c2not(inp0, sum0, anc0);
cnot(inp0, anc0);
```

*Figure 50 –GenAdder's proprietary output for 2-qubit wide adder circuit*

### 7.1.2   Making Genadder QIS-XML compatible

In order to use the GenAdder with QIS-XML, the only necessary change was to adjust the existing program and output the circuit description in XML instead of the proprietary format. Given that the C source code was available and in the public domain, I simply downloaded it from the NIST web site and made the relevant changes. The result is a QIS-XML compliant version of GenAdder that I used to generate adder circuit of various sizes from 2 to 50 bits (6 to 150 qubits). The 2-qubit example in QIS-XML is shown below.

```
<?xml version="1.0" encoding="UTF-8"?>
<c:CircuitLibrary xmlns:c="qis:circuit:1_0"  xmlns:r="qis:reusable:1_0"
xmlns:xsi="http://www.w3.org/2001/XMLSchema-instance">
    <r:Identification>
        <r:ID>genadder</r:ID>
    </r:Identification>
    <c:Circuit size="6">
        <r:Input qubit="1"><r:Name>InputA0</r:Name></r:Input>
        <r:Input qubit="2"><r:Name>InputB0</r:Name></r:Input>
        <r:Input qubit="3"><r:Name>Ancillary0</r:Name></r:Input>
        <r:Input qubit="4"><r:Name>InputA1</r:Name></r:Input>
        <r:Input qubit="5"><r:Name>InputB1</r:Name></r:Input>
        <r:Input qubit="6"><r:Name>Ancillary1</r:Name></r:Input>
        <r:Output qubit="2"><r:Name>Sum0</r:Name></r:Output>
        <r:Output qubit="5"><r:Name>Sum1</r:Name></r:Output>
        <r:Output qubit="6"><r:Name>CarryOut</r:Name></r:Output>
        <!-- sum bit 0 -->
```

```xml
    <c:Step>
        <c:Operation>
            <c:Map qubit="1" input="1"/>
            <c:Map qubit="2" input="2"/>
            <c:Map qubit="3" input="3"/>
            <c:GateRef>
                <r:ID>TOFFOLI</r:ID>
            </c:GateRef>
        </c:Operation>
    </c:Step>
    <c:Step>
        <c:Operation>
            <c:Map qubit="1" input="1"/>
            <c:Map qubit="2" input="2"/>
            <c:GateRef>
                <r:ID>C-NOT</r:ID>
            </c:GateRef>
        </c:Operation>
    </c:Step>
    <!-- sum bit 1 -->
    <c:Step>
        <c:Operation>
            <c:Map qubit="4" input="1"/>
            <c:Map qubit="5" input="2"/>
            <c:Map qubit="6" input="3"/>
            <c:GateRef>
                <r:ID>TOFFOLI</r:ID>
            </c:GateRef>
        </c:Operation>
    </c:Step>
    <c:Step>
        <c:Operation>
            <c:Map qubit="4" input="1"/>
            <c:Map qubit="5" input="2"/>
            <c:GateRef>
                <r:ID>C-NOT</r:ID>
            </c:GateRef>
        </c:Operation>
    </c:Step>
    <c:Step>
        <c:Operation>
            <c:Map qubit="3" input="1"/>
            <c:Map qubit="5" input="2"/>
            <c:Map qubit="6" input="3"/>
            <c:GateRef>
                <r:ID>TOFFOLI</r:ID>
            </c:GateRef>
        </c:Operation>
    </c:Step>
    <!-- finish bit 1 -->
    <c:Step>
        <c:Operation>
            <c:Map qubit="3" input="1"/>
            <c:Map qubit="5" input="2"/>
            <c:GateRef>
                <r:ID>C-NOT</r:ID>
            </c:GateRef>
        </c:Operation>
    </c:Step>
    <!-- finish bit 0 -->
    <c:Step>
```

```
        <c:Operation>
            <c:Map qubit="1" input="1"/>
            <c:Map qubit="2" input="2"/>
            <c:Map qubit="3" input="3"/>
            <c:GateRef>
                <r:ID>TOFFOLI</r:ID>
            </c:GateRef>
        </c:Operation>
    </c:Step>
    <c:Step>
        <c:Operation>
            <c:Map qubit="1" input="1"/>
            <c:Map qubit="3" input="2"/>
            <c:GateRef>
                <r:ID>C-NOT</r:ID>
            </c:GateRef>
        </c:Operation>
    </c:Step>
  </c:Circuit>
</c:CircuitLibrary>
```

*Figure 51 – 2-qubit wide adder circuit in QIS-XML*

### 7.1.3   Circuit Visualization

To visualize the quantum circuit generated by the QIS-XML version of GenAdder, I used an upgraded transforms to convert the XML into SVG. The output for a 3-qubit adder is illustrated below.

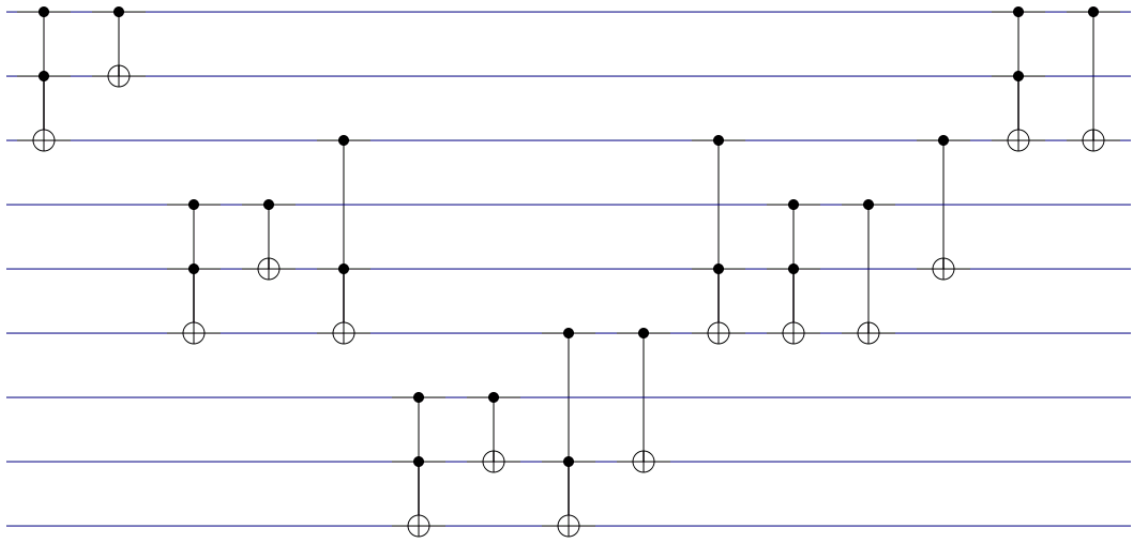

*Figure 52 – SVG rendering of the 3-qubit QIS-XML adder circuit*

## 7.2   INTEGRATING THE CIRCUIT INTO A PROGRAM

Having now a set of quantum adder circuits, using them in a QIS-XML Program is a fairly simple process. For example, creating a program that performs the operation 2+1=3 requires a 2-bit adder that uses 6-qubits. I therefore need a program that declares a 6-qubit <Memory> and executes the 2-qubit adder circuit with a registry initialized to the proper value to represent 2+1 (this is the example presented the previous section to illustrate the new program module in QIS-XML). A slightly more complex example performing the addition 6+7 = 13 is presented below:

```
<!-- 5 qubit adder 6+7 -->
<p:Program>
    <r:Identification>
        <r:ID>six_plus_seven</r:ID>
    </r:Identification>
    <p:Name>Six plus Seven</p:Name>
    <p:Memory size="15"/>
    <p:Execute>
        <p:Register size="15">
```

```
        <p:Prepare>
            <p:QubitSet>
                <!-- A = 6 -->
                <p:QubitIndex>4</p:QubitIndex>
                <p:QubitIndex>7</p:QubitIndex>
                <!-- B = 7 -->
                <p:QubitIndex>2</p:QubitIndex>
                <p:QubitIndex>5</p:QubitIndex>
                <p:QubitIndex>8</p:QubitIndex>
                <p:Value r="1"/>
            </p:QubitSet>
        </p:Prepare>
    </p:Register>
    <p:CircuitRef>
        <r:ID>adder5</r:ID>
    </p:CircuitRef>
</p:Execute>
<p:Measure>
    <p:Register size="6">
        <p:QubitIndex>2</p:QubitIndex>
        <p:QubitIndex>5</p:QubitIndex>
        <p:QubitIndex>8</p:QubitIndex>
        <p:QubitIndex>11</p:QubitIndex>
        <p:QubitIndex>14</p:QubitIndex>
        <p:QubitIndex>15</p:QubitIndex>
    </p:Register>
</p:Measure>
</p:Program>
```

*Figure 53 – A 5-qubit adder QIS-XML program to compute 6+7=13*

The important element here is that we have a complete description of the quantum circuit and its underlying gate in a XML format. This is sufficient information to convert or "compile" this information into other proprietary languages that can be understood by quantum hardware or simulators. Such example is illustrated in the next section.

## 7.3  USING QIS-XML WITH THE QUANTUM COMPUTER LANGUAGE (QCL)

### 7.3.1  Overview

The Quantum Computer Language outlined in 2000 by Bernhard Omer (Omer, Quantum Programming in QCL 2000) is one of the most commonly cited languages for quantum computing. QCL is not only a specification but also comes with a fully operational simulator that can be downloaded from the author's web site.

This simulator basically executes code written in the QCL language. It works under Posix complaint systems and the latest version is QCL 0.6.3 (released in December 2006). It can be used in interactive mode or using external QCL scripts. The software binaries are only available for Linux (but source code is available). For technical reason, the latest version could not be used in my Linux environment and I used release 0.5.1 instead.

### 7.3.2  QIS-XML Compiler

The prototype QIS-XML compiler for QCL does not implement all the gates and features but currently is sufficient to demonstrate the quantum adder circuit example. One potential improvement would be to define each circuit used by a program as a new QCL function, which would reduce the complexity of the generated code (though it wouldn't make a difference in the adder case since the circuit is only used once). The <Register> element implementation is also currently very limited and by default addresses the whole <Memory>.

The "compiler" is our case is simple and standard XSL transformation that turns QIS-XML into QCL code. The resulting 2-qubit adder program in QCL language is shown below. The cryptic names for the quantum register are necessary, as QCL does not allow the global redefinition of variables.

```
// ============================
// QIS-XML QCL Compiler v2007.04
// ============================

int i;
int value;
// Allocate program memory
qureg memory[6];

// WARNING: p:Register not fully implemented ***
qureg registerIDAQ3MTE = memory;
// PREPARE
i = 1;
measure registerIDAQ3MTE[1],value;
if value != 1 { X(registerIDAQ3MTE[1]); }
measure registerIDAQ3MTE[3],value;
if value != 1 { X(registerIDAQ3MTE[3]); }

// CIRCUIT adder2
// STEP 1
// OPERATION 1
qureg registerIDAGFMTE =
registerIDAQ3MTE[0]®isterIDAQ3MTE[1]®isterIDAQ3MTE[2];
CNot(registerIDAGFMTE[2] , registerIDAGFMTE[0] & registerIDAGFMTE[1]);

// STEP 2
// OPERATION 1
qureg registerIDATFMTE = registerIDAQ3MTE[0]®isterIDAQ3MTE[1];
CNot(registerIDATFMTE[1],registerIDATFMTE[0]);

// STEP 3
// OPERATION 1
qureg registerIDA3FMTE =
registerIDAQ3MTE[3]®isterIDAQ3MTE[4]®isterIDAQ3MTE[5];
CNot(registerIDA3FMTE[2] , registerIDA3FMTE[0] & registerIDA3FMTE[1]);

// STEP 4
// OPERATION 1
qureg registerIDAKGMTE = registerIDAQ3MTE[3]®isterIDAQ3MTE[4];
CNot(registerIDAKGMTE[1],registerIDAKGMTE[0]);

// STEP 5
// OPERATION 1
qureg registerIDAUGMTE =
registerIDAQ3MTE[2]®isterIDAQ3MTE[4]®isterIDAQ3MTE[5];
CNot(registerIDAUGMTE[2] , registerIDAUGMTE[0] & registerIDAUGMTE[1]);
```

```
// STEP 6
// OPERATION 1
qureg registerIDABHMTE = registerIDAQ3MTE[2]®isterIDAQ3MTE[4];
CNot(registerIDABHMTE[1],registerIDABHMTE[0]);

// STEP 7
// OPERATION 1
qureg registerIDALHMTE =
registerIDAQ3MTE[0]®isterIDAQ3MTE[1]®isterIDAQ3MTE[2];
CNot(registerIDALHMTE[2] , registerIDALHMTE[0] & registerIDALHMTE[1]);

// STEP 8
// OPERATION 1
qureg registerIDAYHMTE = registerIDAQ3MTE[0]®isterIDAQ3MTE[2];
CNot(registerIDAYHMTE[1],registerIDAYHMTE[0]);

// MEASUREMENT
for i=0 to 5{
    measure memory[i],value;
    print i,"=",value;
}
```

*Figure 54 – QIS-XML generated source code for quantum adder in QCL language*

### 7.3.3   Results

Executing the above 2+1 code in QCL produces the following output. The result (11 in binary) is read on qubit #4 and #1 respectively.

```
QCL Quantum Computation Language (32 qubits, seed 1177862942)
: 0 = 0
: 1 = 1
: 2 = 0
: 3 = 1
: 4 = 1
: 5 = 0
```

*Figure 55 - – QCL simulator output for 2+1 = 3 circuit*

The output of a more complex 5-bit adder (15-qubit) circuit computing 6+7 = 13 is shown below. The result (01101 in binary) is read on qubit 13,10,7,4,1 (other qubits can be ignored except 14 who would be the carry out).

```
QCL Quantum Computation Language (32 qubits, seed 1177862811)
: 0 = 0
: 1 = 1
: 2 = 0
: 3 = 1
: 4 = 0
: 5 = 0
: 6 = 1
: 7 = 1
: 8 = 0
: 9 = 0
: 10 = 1
: 11 = 0
: 12 = 0
: 13 = 0
: 14 = 0
```

*Figure 56 – QCL simulator output for 6+7 = 13 circuit*

### 7.3.4 Summary

QCL is an excellent programming environment with a well-defined language. It is a hybrid language that mixes classic and quantum instructions. While I do not believe such language can take full advantage of quantum algorithm, it is very accessible to classic programmers and an excellent learning and experimenting tool. It should also support many of the features available though QIS-XML and comes with a very flexible quantum memory management model. This probably remains one of the best simulators available today (along with QuiDDPro).

# 8 OTHER POTENTIAL APPLICATIONS

This project has presented some of potential uses of the XML technology applied to QIS. Several other applications are possible as illustrated by the diagram below. The main advantage of using XML is that it comes with a wide range of tools built-in. This greatly reduces the development efforts and ensures compatibility across the frameworks. A few additional ideas and suggestions are presented below.

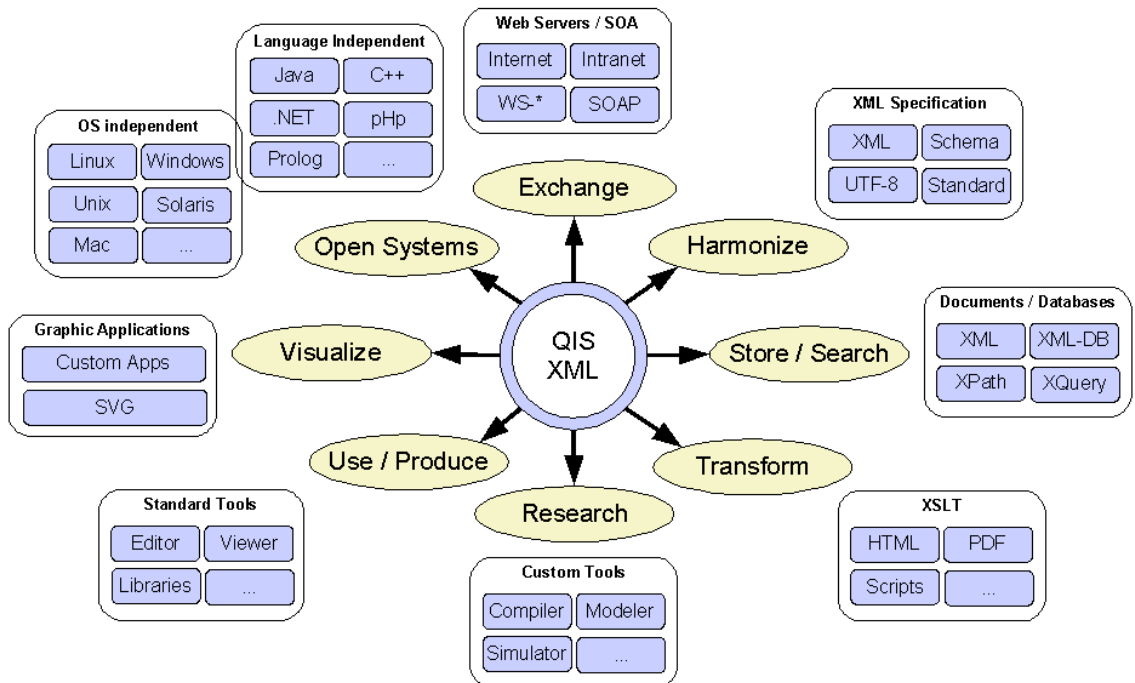

*Figure 57 – The QIS-XML framework*

## 8.1 HARMONIZATION

By using an XML Schema, we have seen that we can outline a model of a well-defined information structure. Making the QIS-XML schema a widely used specification would harmonize the way the QIS community captures, stores and exchanges information about quantum information systems. Defining the information structure is however only a small step towards achieving such a goal. Wide scale acceptance of a specification requires strong endorsement by leading agencies in QIS in order to transform it into a recommended practice. Such initiative could be under the governance of agencies with strong interest in quantum computing, possibly through a membership-based organization (as it is often the case in IT), encompassing both the quantum research and industry IT sectors. Drawing interest from standard governing organizations such as the Institute of Electrical and Electronics Engineers (IEEE), World Wide Web Consortium (W3C) or the Organization for the Advancement of Structured Information Standards (OASIS) could be a sound strategy to achieve this while at the same time foster collaboration with the broad community. Starting a new committee under OASIS for example - a fairly simple process - could potentially be very effective and reach out to a wide audience.

## 8.2 DATABASES AND GLOBAL REGISTRIES

We have seen that with a query language such as XPath, an XML document can in essence be used a database. Collection of XML documents can in turn also be stored in what are called native XML databases (such as BaseX, eXist, Tamino) or hybrid databases (like IBM DB2, Oracle, Microsoft SQL Serve). This enables the use of the Xquery language to perform complex queries across the collection. Using such approach would allow establishing large repositories or knowledge bases of quantum gates, circuits, algorithms and standard XSL transformations that could be made accessible to the broad QIS community. By combining with other XML technologies such as web services, we could also implement what are known as public or private metadata registries to globally share QIS knowledge.

## 8.3   BUILDING TOOLS ON THE STANDARD

Having a standard structure to represent gates, circuits or algorithms, would not only allow to build better tools but also to ensure that such tools would be able to exchange information in a standard format and therefore be compatible with each other. Metadata management software and experimental applications using the QIS-XML could then be made to understand each other.

For example, a package that specializes in circuit design could provide it's input to visualization software or publish directly in a public catalog. A QIS-XML language editor could send its "code" to a remote quantum computer where a QIS-XML compiler could turn the instructions into circuits for execution. Educational software could access public QIS-XML knowledge bases or registries to provide resources to students.

# 9  CONCLUSIONS

Through this research project, I have demonstrated how the XML language and technologies can be used to build an information structure for the capture, storage, research, exchange and processing of quantum gate and circuits metadata. The framework also provides for the representation of this information in a web friendly and user-friendly format.

I have also discussed how the QIS-XML specification can be used to build innovative tools for research or the realization of quantum computer systems, and how the technology can facilitate the publication and exchange of information across the QIS community or with any interested user. Adopting a technology that is widely used and understood by the classic ICT community would foster the better understanding and acceptance of quantum information science by the ICT experts. It will also facilitate the interoperability of classic and quantum information systems. Other activities that would foster a broad adoption of a QIS-XML as a standard metadata structure may also be undertaken.

I hope that this initial draft of QIS-XML will inspire others to build QIS tools around the specification or leveraging XML technology, and to help bridge the gaps that currently exist between the Classic Information Science and Quantum Information Science communities.

I believe that the success of QIS will be closely related to its ability to integrate into the existing CIS framework. This is a challenge that I like to refer to as the "Information Science Grand Unification Challenge" and that could ultimately transform CIS and QIS into a "Complete Information Science".

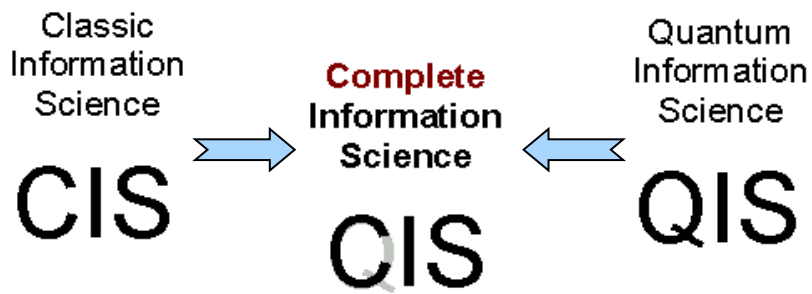

*Figure 58 – The Information Science Grand Unification Challenge*

# ANNEX 1: QIS-XML GENADDER SOURCE CODE

```
/**
 * Generates a quantum circuit in the QIS-XML format
 * to add two binary numbers of any width
 *
 * Author : Pascal Heus
 *
 * Based on the "genadder" utility by Paul E. Black (paul.black@nist.gov)
 * National Institute of Standards and Technology, 2003
 * http://hissa.nist.gov/~black/Quantum/genadder.html
 *
 *
 *
 * This program is free software; you can redistribute it
 * and/or modify it under the terms of the GNU Lesser General Public
 * License as published by the Free Software Foundation; either version
 * 2.1 of the License, or (at your option) any later version.
 *
 * This program is distributed in the hope that it will be useful,
 * but WITHOUT ANY WARRANTY; without even the implied warranty of
 * MERCHANTABILITY or FITNESS FOR A PARTICULAR PURPOSE.
 * See the GNU Lesser General Public License for more details.
 *
 * The full text of the license is available at
 * http://www.gnu.org/copyleft/lesser.html
 *
 */

using namespace std;

#include <iostream>
#include <stdio.h>
#include <stdlib.h> // for exit()
#include <string.h>

#define VERSION "2007.01"
#define USAGE   "Usage: genadder_xml <number_of_bits>"

//-------------------------------------------------------------------------
// Produce top-level gates
// They are factored here so each top-level gate can generate a series of
// lower-level gates, e.g., Steane coding
//-------------------------------------------------------------------------

static void gate_cnot(FILE *p_outfp, int p_control, int p_target)
{
    // OPTIMIZATION I: no-op if control is < 0
```

```
    if (p_control < 0) return;

    fprintf(p_outfp, "\t\t<c:Step>\n");
        fprintf(p_outfp, "\t\t\t<c:Operation>\n");
            fprintf(p_outfp, "\t\t\t\t<c:Map qubit=\"%d\"
input=\"1\"/>\n",p_control+1); // QIS-XML uses 1-based qubit reference
            fprintf(p_outfp, "\t\t\t\t<c:Map qubit=\"%d\"
input=\"2\"/>\n",p_target+1); // QIS-XML uses 1-based qubit reference
            fprintf(p_outfp, "\t\t\t\t<c:GateRef>\n");
            fprintf(p_outfp, "\t\t\t\t\t<r:ID>C-NOT</r:ID>\n");
            fprintf(p_outfp, "\t\t\t\t</c:GateRef>\n");
        fprintf(p_outfp, "\t\t\t</c:Operation>\n");
    fprintf(p_outfp, "\t\t</c:Step>\n");
}

static void gate_c2not(FILE *p_outfp, int p_ctrl1, int p_ctrl2, int p_target)
{
    // OPTIMIZATION I: no-op if control is constant 0
    if (p_ctrl1 < 0  || p_ctrl2 < 0) return;

    fprintf(p_outfp, "\t\t<c:Step>\n");
        fprintf(p_outfp, "\t\t\t<c:Operation>\n");
            fprintf(p_outfp, "\t\t\t\t<c:Map qubit=\"%d\"
input=\"1\"/>\n",p_ctrl1+1); // QIS-XML uses 1-based qubit reference
            fprintf(p_outfp, "\t\t\t\t<c:Map qubit=\"%d\"
input=\"2\"/>\n",p_ctrl2+1); // QIS-XML uses 1-based qubit reference
            fprintf(p_outfp, "\t\t\t\t<c:Map qubit=\"%d\"
input=\"3\"/>\n",p_target+1); // QIS-XML uses 1-based qubit reference
            fprintf(p_outfp, "\t\t\t\t<c:GateRef>\n");
            fprintf(p_outfp, "\t\t\t\t\t<r:ID>TOFFOLI</r:ID>\n");
            fprintf(p_outfp, "\t\t\t\t</c:GateRef>\n");
        fprintf(p_outfp, "\t\t\t</c:Operation>\n");
    fprintf(p_outfp, "\t\t</c:Step>\n");
}

/**
 * Includes an external XML file in the output file
 */
static void includeXmlFile(FILE *p_outfp, char * p_XmlFilename) {
    FILE * pXmlFile;
    char buf[2048];

    pXmlFile = fopen(p_XmlFilename,"rt");
    if(pXmlFile != NULL) {
        while(fgets(buf,sizeof(buf),pXmlFile)) {
            fprintf(p_outfp, "%s",buf);
        }
    }
    else {
        fprintf(p_outfp,"<!-- INCLUDE ERROR: file %s not found --
>\n",p_XmlFilename);
    }
}

//-------------------------------------------------------------------------
// Generate the adder
//
// Bits are zero-based, 0 is least significant.  That is
//      a0, a1, a2, ..., an
//  represents the value
```

```
//       an*2^n + ... + a2*2^2 + a1*2^1 + a0*2^0
// Input is inpN and sumN.  The work qubits are ancN.
// Output is sumN and carryOut.
//
// This follows Jozef Gruska "Quantum Computing", page 96
//       inp == a, sum == b
// Detailed implementation of blocks by Paul E. Black 11 July 2002
// using Generate http://hissa.nist.gov/~black/Quantum/generate.html
//---------------------------------------------------------------------------

/**
 * Generates and adder circuit for specific bit
 */
static void generateAdderGates(FILE *p_outfp, int p_numBits, int p_currentBit)
{
    // sum for this stage
    fprintf(p_outfp,"\t\t<!-- sum bit %d -->\n", p_currentBit);
    gate_c2not(p_outfp, (p_currentBit*3)+0, (p_currentBit*3)+1,
(p_currentBit*3)+2);
    gate_cnot( p_outfp, (p_currentBit*3)+0, (p_currentBit*3)+1);
    gate_c2not(p_outfp, (p_currentBit*3)-1, (p_currentBit*3)+1,
(p_currentBit*3)+2);

    if (p_numBits > 1) {
        // create the rest of the stages (recursive)
        generateAdderGates(p_outfp, p_numBits-1, p_currentBit+1);
    }

    // finish this stage - undo ancilla and form sum w/carry-in
    if (p_numBits > 1 || p_currentBit > 0) {
        fprintf(p_outfp,"\t\t<!-- finish bit %d -->\n", p_currentBit);
    }
    if (p_numBits > 1) {
        gate_c2not(p_outfp, (p_currentBit*3)-1, (p_currentBit*3)+1,
(p_currentBit*3)+2);
        gate_c2not(p_outfp, (p_currentBit*3)+0, (p_currentBit*3)+1,
(p_currentBit*3)+2);
        gate_cnot( p_outfp, (p_currentBit*3)+0, (p_currentBit*3)+2);
    }
    gate_cnot(p_outfp, (p_currentBit*3)-1, (p_currentBit*3)+1);
}

/*
 * Generates the adder
 */
static void generateAdderCircuit(FILE *p_outfp, int p_numBits)
{
    int i;

    // Initialize XML
    fprintf(p_outfp, "<?xml version=\"1.0\" encoding=\"UTF-8\"?>\n");
    fprintf(p_outfp, "<c:CircuitLibrary xmlns:c=\"qis:circuit:1_0\"
xmlns:r=\"qis:reusable:1_0\" xmlns:xsi=\"http://www.w3.org/2001/XMLSchema-
instance\">\n");

    fprintf(p_outfp, "\t<r:Identification>\n");
        fprintf(p_outfp, "\t\t<r:ID>genadder</r:ID>\n");
    fprintf(p_outfp, "\t</r:Identification>\n");
```

```
    // Start adder circuit
    fprintf(p_outfp, "\t<c:Circuit size=\"%d\">\n",p_numBits*3);

    // Label the inputs
    for(i=0 ; i < p_numBits ; i++) {
        fprintf(p_outfp, "\t\t<r:Input
qubit=\"%d\"><r:Name>InputA%d</r:Name></r:Input>\n",(i*3)+0+1,i); // QIS-XML
uses 1-based qubit reference
        fprintf(p_outfp, "\t\t<r:Input
qubit=\"%d\"><r:Name>InputB%d</r:Name></r:Input>\n",(i*3)+1+1,i); // QIS-XML
uses 1-based qubit reference
        fprintf(p_outfp, "\t\t<r:Input
qubit=\"%d\"><r:Name>Ancillary%d</r:Name></r:Input>\n",(i*3)+2+1,i); // QIS-XML
uses 1-based qubit reference
    }
    // Label the outputs
    for(i=0 ; i < p_numBits ; i++) {
        fprintf(p_outfp, "\t\t<r:Output
qubit=\"%d\"><r:Name>Sum%d</r:Name></r:Output>\n",(i*3)+2,i); // QIS-XML uses
1-based qubit reference
    }
    fprintf(p_outfp, "\t\t<r:Output
qubit=\"%d\"><r:Name>CarryOut</r:Name></r:Output>\n",(p_numBits*3)-1+1); //
QIS-XML uses 1-based qubit reference

    // Generate all steps
    generateAdderGates(p_outfp, p_numBits, 0);

    // End circuit
    fprintf(p_outfp, "\t</c:Circuit>\n");

    // End XML
    fprintf(p_outfp, "</c:CircuitLibrary>\n");
}

/**
 * MAIN
 */
int main(int argc, char *argv[])
{
    // Check arguments
    if (argc > 1 && strcmp(argv[1], "--version") == 0) {
        cout << "genadder_xml Version " << VERSION << endl;
        exit(0);
    }

    if (argc != 2) {
        cout << "Wrong number of operands" << endl;
        cout << USAGE << endl;
        exit(-1);
    }

    // Get the number of bits
    int numBits = 0;

    char dummy;

    if (sscanf(argv[1], " %d%c", &numBits, &dummy) != 1) {
        cout << "Invalid number of bits: " << argv[1] << endl;
```

```cpp
        cout << USAGE << endl;
        exit(-1);
    }

    if (numBits < 1) {
        cout << "Number of bits must be positive: " << numBits << endl;
        cout << USAGE << endl;
        exit(-1);
    }

    // Generate the circuit
    generateAdderCircuit(stdout, numBits);

    return 0;
}
```

# REFERENCES

# CURRICULUM VITAE

Pascal Heus graduated in computer science in 1985 from the Institut Paul Lambin at the Catholic University of Louvain in Belgium. He resumed his studies in 2000 and earned a graduate Certificate in Computational Technique and Applications Georges Mason University, VA in 2003. He later on enrolled in the Master program in 2006, focusing on Quantum Information Science.

Pascal Heus is an Information Technology specialist with over 20 years of professional experience and a strong focus on data and metadata management solutions for the archival and dissemination of socio-economic microdata and official statistics. He worked for many years with the World Bank and the International Household Survey Network (IHSN), leading the development and deployment of the institutional microdata management systems and working closely with national statistical agencies in developing countries towards the establishment of data archives and dissemination solutions. He's an active expert of the DDI Alliance and is closely involved in the development of the DDI XML specification, an international effort to establish a standard for technical documentation describing social science data.

In 2006, Pascal helped establish the Open Data Foundation where he is currently Executive Manager. In September 2007, he joined Metadata Technology where he is now Vice-President for North America, focusing on consultancy, software development, training, and services helping organizations better collect, disseminate, discover, and analyze statistical, scientific and research data and metadata. In 2010, Pascal established Integrated Data Management Services Inc., a company offering data and metadata processing services complementing Metadata Technology's IT solutions